\documentclass{JFM-FLM_Au}
\renewcommand{\oddfooterflagdefns}[1]{\hbox to \textwidth{\hfill\thepage}}
\renewcommand{\evenfooterflagdefns}[1]{\hbox to \textwidth{\thepage\hfill}}

\usepackage{subfig}
\usepackage{float}
\usepackage{subcaption}
\usepackage{graphicx}
\usepackage{epstopdf}

\lefttitle{E.E. Nicotra, F.A. Centlivre, J. Poggie, S.M. Bane, and J.S. Jewell}
\righttitle{Exp. Investigation of Hypersonic SWBLI Dynamics Using Plasma-Generated Disturbances}

\title{Experimental Investigation of Hypersonic Shock-Wave/Boundary-Layer Interaction Dynamics Using Plasma-Generated Disturbances}

\author{Emma E. Nicotra\aff{1}, Francis A. Centlivre\aff{1},
Jonathan Poggie\aff{1}, Sally P.M. Bane\aff{1}, and Joseph S. Jewell\aff{1}}

\affiliation{\aff{1}Purdue University, West Lafayette, IN, USA}

\corresau{Emma E. Nicotra, \email{enicotra@purdue.edu}}

\begin{document}
\maketitle

\begin{abstract}
The unsteady dynamics of a transitional 3D shock-wave/boundary-layer interaction (SWBLI) are investigated using pulsed plasma disturbances on a cone-slice-ramp geometry in the Boeing/AFOSR Mach-6 Quiet Tunnel. Nanosecond-duration repetitively pulsed spark discharges introduce disturbances which convect through the interaction, causing transient turbulent breakdown in the reattaching shear layer. The unsteady behavior of the separation bubble is characterized using schlieren imaging, and natural broadband unsteadiness is observed in the form of large-scale shock and separation flapping at frequencies below 1500~Hz ($Sr_L< 0.25$). Disturbance packets are introduced at a range of perturbation frequencies and generally induce shear layer oscillations; however, perturbation frequencies below the natural cutoff frequency ($f\approx 1500$ $\mathrm{Hz}$) yield fluctuations with significantly larger amplitudes than those above this frequency. Phase-averaged schlieren images reveal that the disturbances cause a decrease in shear layer height immediately after inducing turbulent breakdown around the reattachment point, with this displacement beginning downstream and moving upstream. The shear layer's response to this displacement is consistent with that of an under-damped dynamic system with a preferred oscillation frequency; the measured frequency of the oscillatory response to external perturbation ($f = 1464$ Hz) is also the approximate cutoff frequency of the natural broadband unsteadiness. Furthermore, the relationship between perturbation frequency and oscillation amplitude resembles that of a low-pass filter, with a cutoff frequency that is within 5\% of the measured oscillator frequency. These experimental findings support the theory that the cutoff frequency for the SWBLI's frequency-selective amplification is tied to the time scale of the weakly damped global mode.  
\end{abstract}

\section{Introduction}

Separation bubbles resulting from shock-wave/boundary-layer interactions (SWBLI) often exhibit significant unsteadiness which introduces oscillatory pressure loads, leading to aero-structural fatigue and decreased inlet and control surface performance \citep{clemens_low-frequency_2014}. One of the most significant facets of SWBLI-induced unsteadiness is the large-scale, low-frequency oscillation of the separation bubble and shock, but the exact mechanism responsible for this behavior is still debated \citep{dolling_fifty_2001,clemens_low-frequency_2014,gaitonde_dynamics_2023}. The origins of the unsteadiness have been the topic of frequent experimental and computational investigation, with specific interest and conflicting results around its sensitivity to upstream and downstream mechanisms \citep{dupont_space_2006,clemens_low-frequency_2014,adler_dynamic_2018}.  

\subsection{Origins of Separation Unsteadiness}
Studies on SWBLI-induced unsteadiness are most often conducted on turbulent interactions \citep{dolling_fifty_2001,clemens_low-frequency_2014,gaitonde_dynamics_2023}. \cite{plotkin_shock_1975} developed a mathematical model of the unsteadiness in which the shock wave is convected from its mean location by velocity fluctuations in the turbulent incoming boundary layer and recovers its position via a linear restoration mechanism. 
Compression-corner experiments by \cite{beresh_relationship_2002} and \cite{ganapathisubramani_low-frequency_2009} observed a significant correlation between separation location and near-wall streamwise-velocity fluctuations in the turbulent boundary layer. Simulations by \cite{porter_selective_2019} on a supersonic compression corner found the same relationship by taking conditional averages of the boundary layer velocity profiles based on shock location. However, other studies suggest that alternative, downstream mechanisms are responsible for the SWBLI-induced unsteadiness. \cite{wu_analysis_2008} performed direct numerical simulations (DNS) on a turbulent interaction and found that the shock and separation motion were correlated with and preceded by motion at the reattachment point. Based on this finding, they suggested that the unsteadiness could be explained by a feedback loop between the separation bubble, shear layer and shock. \cite{piponniau_simple_2009} suggested that the phenomenon is caused by a cycle of fluid entrainment by the mixing layer and the subsequent recovery of mass within the bubble. Their model proposed that the shedding of vortices generated in the turbulent mixing layer entrains fluid from the separation region, and that the flapping of the downstream shear layer allows mass flux into the bubble to restore what has been depleted. Based on DNS of a turbulent SWBLI, \cite{priebe_low-frequency_2012} observed a low-frequency modulation of energetic turbulent structures in the shear layer, which they related to a modulation of fluid entrainment rate as a potential origin of the unsteadiness. They proposed that the unsteadiness is due to a downstream global instability in the separation region, with the upstream boundary layer having a weak influence over fully separated flows. 

Using the Navier-Stokes equations, \cite{touber_low-order_2011} derived a governing equation for the shock motion which was mathematically identical to that of Plotkin and agreed well with experimental and large eddy simulation (LES) data. Their analytical model posited that the SWBLI acts as a frequency-selective amplifier for external disturbances, amplifying existing fluctuations up to a certain cutoff frequency. They argued that the broadband low-frequency unsteadiness does not originate from specific low-frequency forcing but rather from the inherent dynamics of the SWBLI system. Based on the work of \cite{touber_low-order_2011}, and in an attempt to reconcile the various theories, \cite{clemens_low-frequency_2014} proposed that the global instability is that of a forced dynamical system which responds to external disturbances from both the upstream boundary layer and downstream shear layer. In this model, the separation bubble has a preferred oscillation frequency inherent to the SWBLI system. Resolvent analysis by \cite{bugeat_low-frequency_2022} on a laminar oblique SBWLI found that the system's transfer function resembles a first-order low-pass filter; their findings suggested that SWBLI-induced unsteadiness was the result of a damped dynamical system under continuous forcing, rather than an intrinsic property of the flow. Similarly, in their detached eddy simulations of a sharp fin, \cite{ngoh_effects_2025} noted that the interaction was only unsteady when supplied with inflow turbulence. However, in simulations of a blunt fin under similar conditions, the interaction showed evidence of self excited oscillations in the absence of inflow turbulence \citep{ngoh_detached_2022}, suggesting that not all interactions can be described by a single dynamic system; rather, they can be classified as a selective amplifier, a self-excited oscillator, or a weakly damped global mode which exhibits characteristics of both.

Many previous studies have studied the dynamic behaviors of turbulent interactions, with far less work on transitional interactions \citep{dolling_exploratory_2003,murphree_experimental_2006,  combs_unsteady_2019,gaitonde_dynamics_2023}. \cite{threadgill_transitional_2021} studied an interaction-transitional compression-corner interaction, and observed low-frequency unsteadiness at a Strouhal number ($Sr_L = fL_{\mathrm{bub}}/U_\infty$) consistent with turbulent SWBLI unsteadiness ($Sr_L \approx 0.025$), along with evidence that a downstream mechanism was driving the unsteadiness. \cite{sansica_forced_2014} conducted 2D DNS on a forced laminar SWBLI and concluded that the low-frequency unsteadiness was most prominently tied to mechanisms inside the separation region. By forcing a laminar interaction in 3D DNS, \cite{sansica_instability_2016} prompted transition near reattachment and linked the presence of low-frequency unsteadiness to upstream-traveling disturbances resulting from turbulent breakdown around reattachment.

\subsection{Control of Shock-Wave/Boundary-Layer Interactions}
Despite uncertainty over the origins of low-frequency SWBLI unsteadiness, there is a consensus that control mechanisms must be developed to address its unwanted effects. \cite{dolling_fifty_2001} published an extensive review of the previous 50 years of SWBLI work in which he emphasized future topics to be explored; these included determining the cause of the low-frequency unsteadiness, further study of three-dimensional interactions, and the development of flow control methods. While acknowledging that attempting flow control when the underlying mechanism is still not fully understood may be considered a futile endeavor by some, Dolling opined that the flow control strategies themselves may be a method for better understanding the system by studying its response to the variables introduced by the control mechanisms. A review of research on low-frequency SWBLI unsteadiness by \cite{clemens_low-frequency_2014} recommended future work using pulsed disturbances in the upstream boundary layer to study the unsteady system dynamics. In a review of 3D SWBLI, \cite{gaitonde_dynamics_2023} state the need for new control approaches, emphasizing that small perturbations are desirable, as they will scale better to larger systems. 

There exists a range of possible SWBLI control objectives, such as decreasing separation length, reducing loads caused by pressure fluctuations, and shifting the frequency of the unsteadiness \citep{dolling_fifty_2001}. Passive control methods, such as shock bumps \citep{bruce_review_2015} and vortex generators \citep{lu_review_2011,pickles_control_2020}, have been tested on various SWBLI configurations with the intention of reducing separation length. Their lack of adaptability, however, limits their performance at off-design conditions. One of the main control mechanisms used in both passive and active strategies is delaying flow separation by introducing vortices to entrain high-momentum fluid into the boundary layer \citep{pickles_control_2020}. Pitched wall jets have been used to generate vortices and inject momentum into the upstream boundary layer, with multiple studies finding a reduction in separation length \citep{souverein_effect_2010,verma_shockwaveboundary-layer_2012,sebastian_control_2023}. \cite{selig_effect_1991} introduced periodic blowing in a turbulent boundary layer upstream of a compression corner and found that small levels of blowing, when introduced at the right location, effectively controlled the frequency of the separation unsteadiness. \cite{narayanaswamy_control_2012} used spark plasma filaments within a cavity to create pulsed wall jets at frequencies on the order of 1~kHz. When these jets were placed upstream of a turbulent SWBLI, the SWBLI unsteadiness became locked with the forcing frequency. Also, the mean shock location moved downstream when forcing at frequencies higher than those associated with the natural unsteadiness.

Surface-mounted plasma actuators have also been investigated as an SWBLI control method. Localized arc-filament plasma actuators (LAFPAs) generate plasma filaments by applying high-voltage pulses ($>$~1~$\upmu$s duration) between two surface-mounted electrodes, introducing local temperature and pressure gradients \citep{utkin_development_2006, caraballo_supersonic_2009}. \cite{yan_investigation_2007} found that the rapid heating creates a ``thermal bump" which acts as an artificial vortex generator, producing a pair of counter-rotating streamwise vortices downstream of the actuator. The prevailing theory is that LAFPAs exert control authority over the flow using the same mechanism as the other control methods discussed, in that they rely on vortex generation to entrain momentum into the boundary layer \citep{caraballo_supersonic_2009,tian_mechanisms_2021}. Because LAFPAs feature a wide operating bandwidth, they have been regarded as a promising method for altering the unsteadiness associated with SBWLI \citep{caraballo_supersonic_2009,webb_control_2013}. \cite{webb_control_2013} used LAFPAs to generate disturbances in a turbulent boundary layer upstream of an impinging shock interaction. They found that the actuator shifted the reflected shock downstream, but they were unable to manipulate the low-frequency unsteadiness. 
Similar to local arc filaments, nanosecond-duration repetitively pulsed (NRP) plasma filaments can be formed between two flush-mounted electrodes; however, as the name suggests, NRP plasmas occur on a significantly shorter time scale. NRP discharges introduce ultrafast heating, with \cite{pai_nanosecond_2010} and \cite{xu_experimental_2011} measuring local temperature increases greater than 1000 K only 30-50~ns after the formation of an NRP discharge at atmospheric pressure. \cite{mckiernan_instability_2020} used nanosecond-duration spark discharges to generate perturbations upstream of a 3D SWBLI in quiet Mach 6 flow; he observed the perturbations causing periods of turbulent breakdown in the downstream shear layer as they convected through the reattachment region. Previous work by \cite{nicotra_plasma_2024} presented preliminary testing of NRP plasma discharges upstream of an interaction-transitional 3D SWBLI on a cone slice ramp and found evidence that the frequency of the unsteady separation motion was locking to the perturbation frequencies. In further experiments, \cite{nicotra_perturbation_2026} perturbed a laminar axisymmetric separation bubble using an NRP plasma discharge; the relationship between forcing frequency and response amplitude resembled that of a low-pass filter. 

\subsection{Cone-Slice-Ramp Configuration}
The cone-slice-ramp geometry consists of a compression ramp mounted on the flat surface of a sliced cone; it generates a complex three-dimensional SWBLI which resembles the effects of a deflected control surface \citep{oberkampf_joint_1992}. Experimental \citep{oberkampf_joint_1992,mckiernan_instability_2021,francis_effect_2024} and computational \citep{walker_joint_1992,thome_computational_2018} studies of the cone-slice-ramp have noted a three-dimensional separation bubble with curved separation and reattachment fronts. Another three-dimensional component of the cone-slice-ramp flow is the pair of counter-rotating vortices inside of the separation bubble, as observed in computations by \cite{thome_computational_2018}. Experimental studies \citep{mckiernan_instability_2020,francis_instability_2024,nicotra_plasma_2024} of the cone-slice-ramp in low-disturbance Mach-6 flow found that the separation bubble and shock exhibited a low-frequency unsteadiness with the separation point fixed at the leading edge of the slice. The flow separation on the cone-slice-ramp differs from separation bubbles in most other SWBLI studies, in that the separation point and shock foot are fixed at the leading edge of the slice, and only the reattachment point is free to move. Based on the definitions provided by \cite{weiss_low-frequency_2021}, this unsteadiness is classified as a \textit{flapping} motion as opposed to the \textit{breathing} motion more commonly observed in SWBLI studies. Although \textit{flapping} is also used to describe the shear layer oscillations near reattachment, in this work it refers to the large-scale motion of both the shear layer and shock.

\subsection{Freestream Noise Effects}
There exist very few studies on the impact of freestream noise on SWBLI unsteadiness \citep{weiss_effect_2007,leidy_tunnel_2021}, with nearly all experimental work performed in facilities with conventional freestream noise \citep{dolling_fifty_2001,bookey_new_2005}. The freestream flow in quiet wind tunnels contain pressure fluctuation levels that are one or more orders of magnitude below those in conventional facilities \citep{schneider_development_2008}. \cite{weiss_effect_2007} investigated freestream noise effects in a quiet supersonic tunnel using a compression ramp with a turbulent upstream boundary layer. While they concluded that the dynamic behavior was not influenced by freestream noise, their low-disturbance core flow ended upstream of the separation shock, and the interaction region was subjected to conventional freestream noise levels. \cite{nicotra_effect_2026} conducted a study of freestream noise effects on separation unsteadiness resulting from an upstream transitional interaction in Mach-6 flow; results showed that greater freestream noise levels caused a significant increase in the magnitude of the unsteady motion and the resulting low-frequency pressure fluctuations inside the bubble. \cite{leidy_tunnel_2021} observed altered trends in separation length and pressure fluctuation spectra when conducting transitional SWBLI studies in a conventional supersonic wind tunnel where the freestream fluctuations triple above a known Reynolds number. They suggested that SWBLI studies in quiet tunnels were vital, as the effects of tunnel noise have never been fully quantified.

\subsection{Present Work}
The objective of this work is to use spark-discharge-generated disturbance packets to selectively perturb an interaction-transitional SWBLI as a way to investigate its dynamic behavior. Experiments are performed in a low-disturbance Mach-6 wind tunnel on a cone-slice-ramp model, with NRP spark discharges introduced to the laminar upstream boundary layer. The goal is to excite the unsteady dynamics via the downstream mechanism by introducing vorticity and prompting turbulent breakdown in the reattaching shear layer. This provides an opportunity to examine the SWBLI system's dynamic response to external perturbation when it is not influenced by significant wind tunnel noise or boundary layer turbulence. The system's receptivity to a range of perturbation frequencies and energies is also investigated. Simulations of select experimental conditions are featured in a companion paper by \cite{centlivre_computational_2026}, which provides further insight into the mechanisms at work in these experiments.

\section{Experimental Methods}

\subsection{Boeing/AFOSR Mach-6 Quiet Tunnel}
\begin{figure}[h]
 \includegraphics[scale=.5]{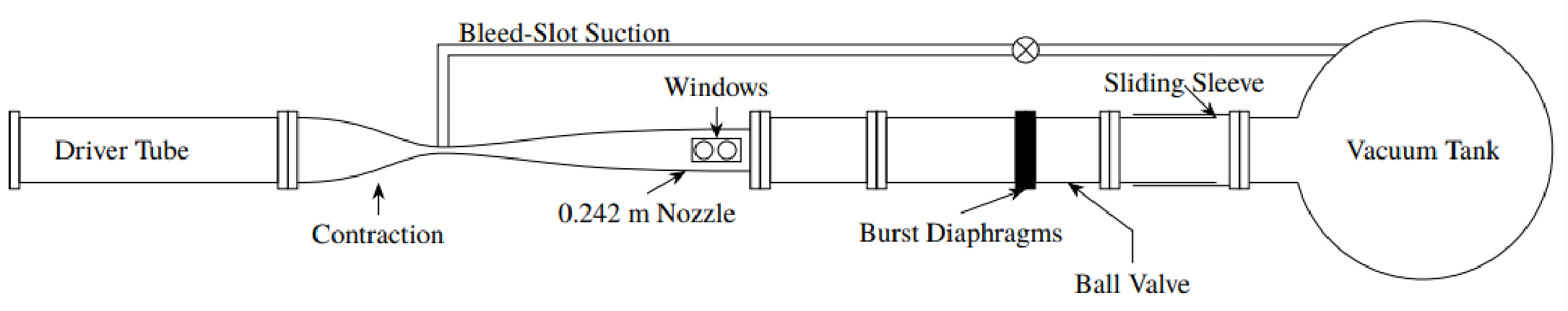}
 \centering
 \caption{Schematic of the Boeing/Air Force Mach 6 Quiet Tunnel (BAM6QT).}
 \label{tunnel}
 \end{figure}

 The Boeing/AFOSR Mach-6 Quiet Tunnel (BAM6QT) at Purdue University is a Ludwieg tube that generates a low-disturbance freestream; a schematic of the BAM6QT is shown in figure \ref{tunnel}. A tunnel test begins by bursting the pair of diaphragms downstream of the test section, allowing a Mach~6 freestream to develop; expansion waves reflect down the length of the driver tube every 200~ms, each causing a decrease in stagnation pressure \citep{gray_flowfield_2019}. These pressure drops allow the model to experience a range of Reynolds numbers throughout a single test, during which quiet flow is maintained for approximately 2~s. The flow in the BAM6QT contains freestream pressure fluctuations on the order of 0.01\%, compared to 0.5 to 3\% in conventional wind tunnels \citep{mamrol_freestream_2021}. This is achieved, in part, by polishing the nozzle to a mirror finish and suctioning the boundary layer at the throat via a bleed valve, allowing a laminar boundary layer to form in the diverging section of the nozzle. The tunnel can also be operated with conventional levels of freestream noise by disabling the boundary-layer suction, resulting in a Mach~5.8 freestream \citep{gray_flowfield_2019}. 
The unit freestream Reynolds number for all experiments was 6.39$\pm0.09\times10^6$~/m. The data presented are from tests performed with the bleed valve open, resulting in a quiet Mach~6 freestream. Stagnation pressure was measured using an ETL-79-HA-DC-190 Kulite pressure sensor installed in the nozzle wall just upstream of the contraction section. The tunnel was filled with heated air so that the stagnation temperature at the start of each test was approximately 433~K.     

A Z-type schlieren system implemented by \cite{price_high-speed_2021} was used to image the motion of the separation bubble and the boundary layer disturbances caused by the plasma discharges. A Cavilux Smart pulsed laser was used to image the separated shear layer over the slice at frequencies up to 90~kHz. The images of the upstream boundary layer were collected using a Cavilux Ultra High Speed pulsed laser at 875~kHz. For both cases, the laser pulse duration was set to 10~ns to prevent blurring of the flow structures. The pulsing was synced to the shutter of a Phantom TMX high speed camera which records the schlieren images.

\subsection{Wind Tunnel Model}
\begin{figure}[h]
 \includegraphics[scale=.4]{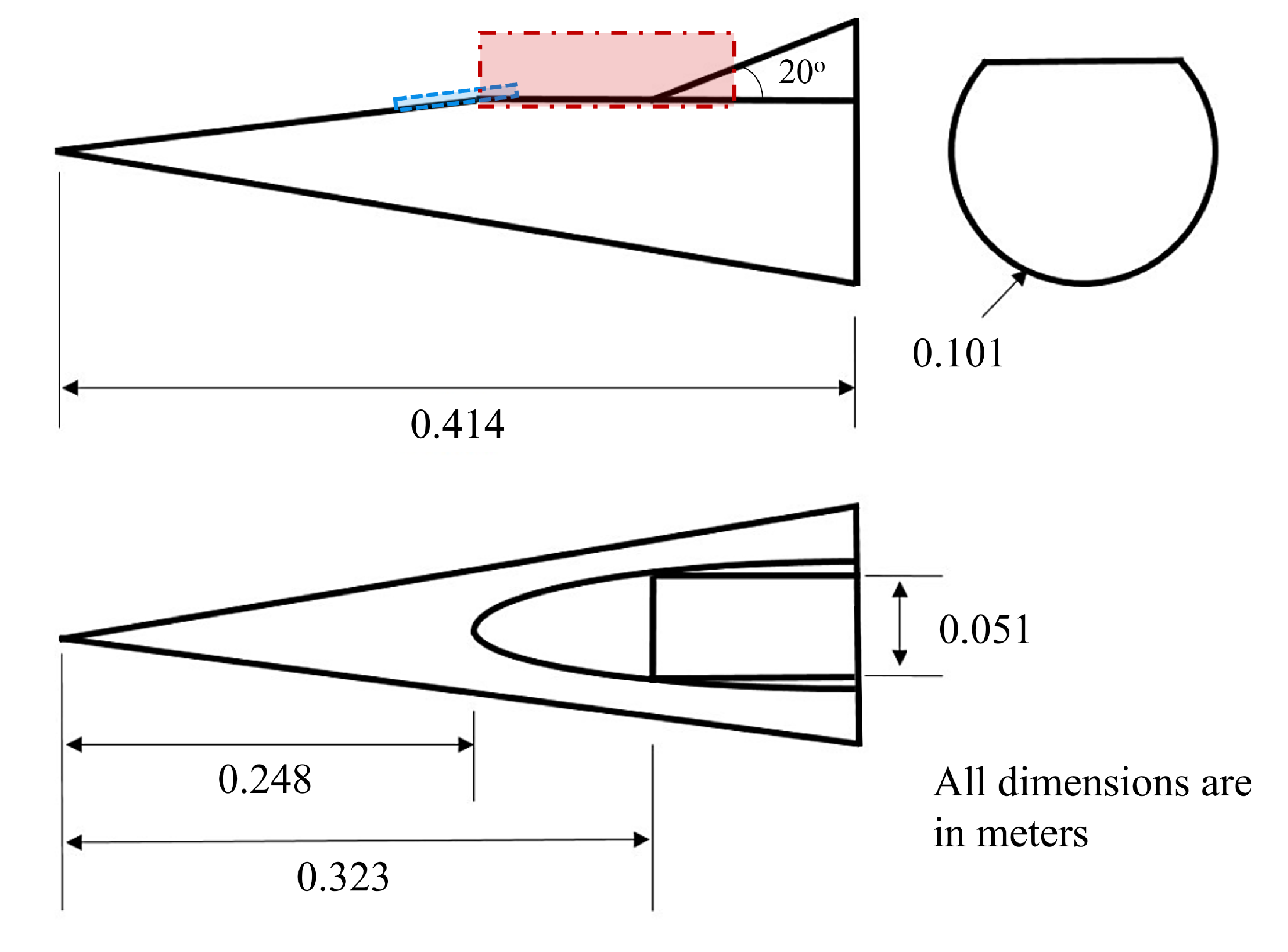}
 \centering
 \caption{Schematic of the cone-slice-ramp wind tunnel model, adapted from \cite{mckiernan_instability_2020}. The red and blue boxes represent the schlieren fields of view for the separation region and upstream boundary layer, respectively.}
 \label{modelDim}
 \end{figure}

The model used in these experiments is a cone-slice-ramp geometry developed by \cite{mckiernan_instability_2020}, as a modified version of the geometry first designed by \cite{oberkampf_joint_1992}. The model, as shown in figure \ref{modelDim}, features a seven-degree half-angle cone with a planar slice taken parallel to the cone's axis of symmetry; a 20-degree ramp is mounted 0.075~m downstream of the leading edge of the slice. A stainless steel nose tip with a tip radius of 116~$\upmu$m was threaded onto the aluminum body of the cone. The cone body was attached to a base plate which connected to a steel adapter that allowed for precise pitch and yaw adjustments. This adapter was connected to a hollow metal sting and mounted along the tunnel centerline. 

\subsubsection{Instrumentation}
The model was instrumented with surface-mounted PCB~132B38 and Kulite~XCE-062 pressure transducers. Four azimuthally-spaced PCB sensors were placed at a fixed axial location upstream of the slice to aid in angle-of-attack adjustments. The model was tested at a nominally zero-degree angle-of-attack, with the precision adapter allowing for corrections in order to align the second-mode frequencies measured by these four sensors, as seen in \cite{gray_measurements_2020}. The angle-of-attack adjustments were performed using tests in noisy flow at $Re_{\infty} \approx$ 9.5~$\times 10^6$~/m. Once aligned, the peak second-mode frequencies were within 2\% of one another, corresponding to an estimated off-axis angle of 0.1$^\circ$, based on the approximation described in \cite{gray_instability_2022}.

 \begin{figure}[h]
 \includegraphics[scale=.25]{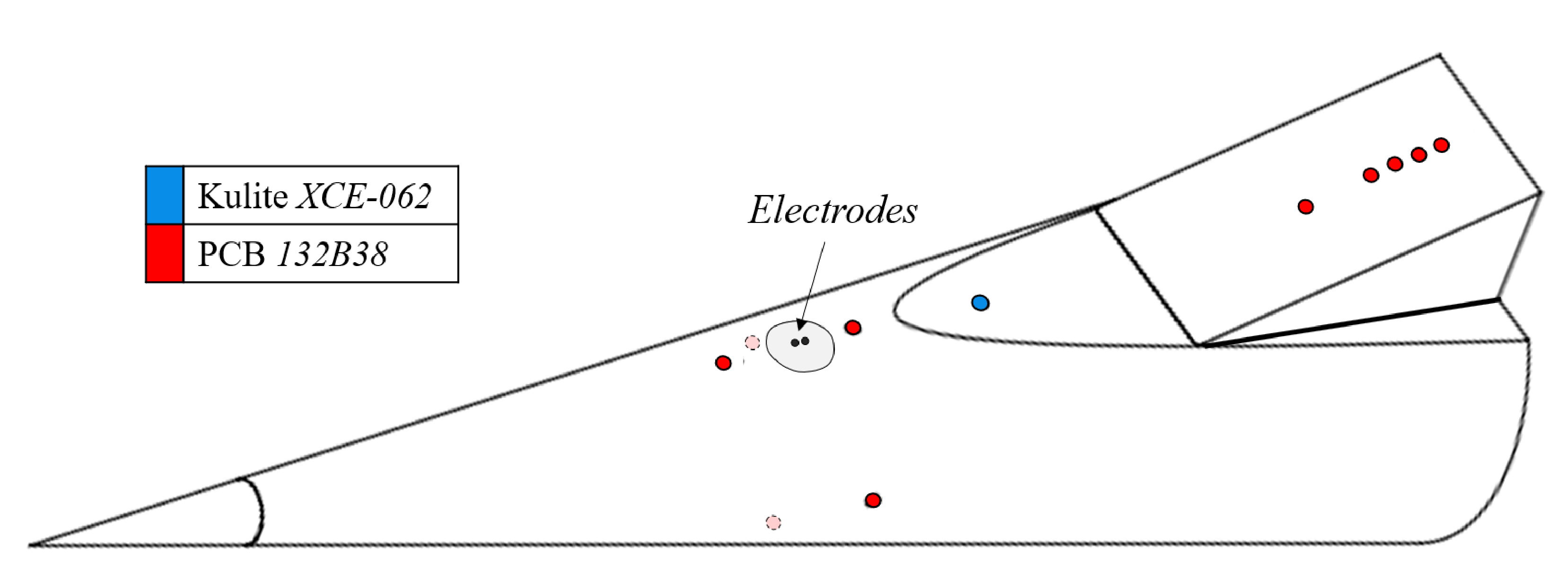}
 \centering
 \caption{Schematic of the cone-slice-ramp wind tunnel model with the plasma actuator and sensor locations identified, adapted from \cite{mckiernan_instability_2020}.}
 \label{model}
 \end{figure} 

\subsubsection{Plasma Actuator}
The NRP plasma actuator consists of two flush-mounted steel  rods housed in an insert made of MACOR, a machinable ceramic material. These electrodes, which can be seen in figure \ref{model}, are aligned so that they are parallel with the flow direction. The rods are 1.588~mm in diameter, and they are installed in the MACOR insert so that, on the surface of the body, they are 0.635~mm apart at their closest point. Each electrode is soldered to a 20-kV-rated wire, which was routed through the hollow metal sting and out through the tunnel wall. The wires and sting were both wrapped in aluminum tape in order to reduce the electromagnetic interference (EMI) produced by the high-voltage pulse generator. The actuator insert is located 221~mm downstream of the nose-tip and approximately 27~mm upstream of the leading edge of the slice. The actuator was originally designed by \cite{mckiernan_instability_2020} to generate both glow and filamentary (``spark") plasma discharges in an attempt to excite boundary layer instabilities in the BAM6QT. In the present work, the plasma discharges were in the spark regime.

\subsection{Plasma Generation}\label{sec:plasma_gen}
\begin{figure}[h]

  \centering
  \subfloat[]{\centering\includegraphics[width=0.45\textwidth]{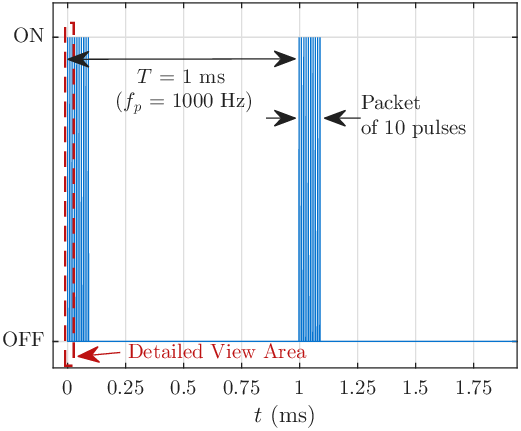}\label{fig:f1}}
  \hfill
  \subfloat[]{\centering\includegraphics[width=0.48\textwidth]{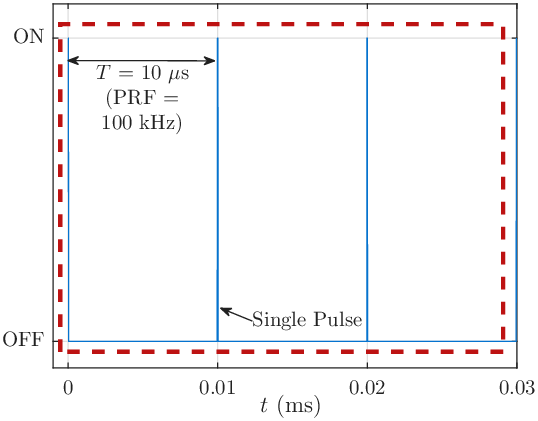}\label{fig:f2}}
  \caption{(a) Sample plasma timing diagram for two packets of 10 pulses each at $f_{p} = 1000$~Hz. (b) Detailed view of the plasma timing for three individual 40~ns duration pulses (PRF = 100~kHz).}
  \label{pulsedia}
\end{figure}

The high-voltage nanosecond-duration pulses applied to the electrodes were generated using an Eagle Harbor Technologies NSP-330-20-F nanosecond pulser. The Eagle Harbor pulser can generate high-voltage pulses with durations between 20 and 110~ns, a peak voltage of 20~kV, and continuous repetition rates up to 100~kHz. Voltage and current measurements were taken by probes connected to the high-voltage output wires and recorded by an Agilent Technologies DS09104A Oscilloscope, which has a maximum capture frequency of 1~GHz. The pulser was triggered using a series of two delay generators to achieve the desired pulse timings. The plasma pulses were applied in high-frequency packets, with a pulse repetition frequency (PRF) of 100~kHz; this frequency was constant throughout all experiments. These packets contained either 5 or 10 pulses each, depending on the case (see table \ref{tab:table1}). The packets were then introduced at a range of packet repetition frequencies, hereafter referred to as the perturbation frequency, $f_p$, ranging from 500 to 5000~Hz. The frequencies tested for all cases were 500, 1000, 1500, 2000, and 5000~Hz. Perturbation frequencies of 1200 and 1335~Hz were added after initial testing to explore behavior between 1000 and 1500~Hz, and they were only tested with Case 4. Figure \ref{pulsedia} shows the timing diagram for a sample test, in which Case~4 packets of 10 pulses are applied at a perturbation frequency of 1000~Hz.

\begin{figure}[h]

  \centering
  \subfloat[]{\includegraphics[width=0.49\textwidth]{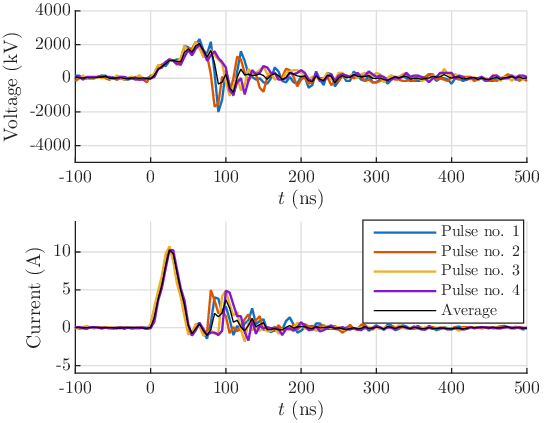}\label{fig:f1}}
 \hfill
  \subfloat[]{\includegraphics[width=0.49\textwidth]{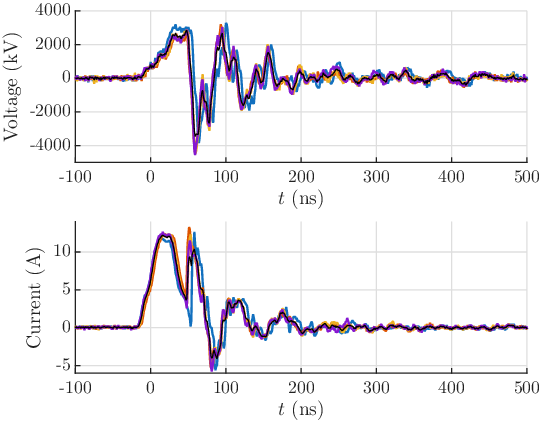}\label{fig:f2}}
  \caption{Sample and average voltage and current traces for NRP plasma filaments with (a) a 20~ns pulse width and (b) a 40~ns pulse width.}
  \label{pulseIV}
\end{figure}

Packets of high-frequency pulses were chosen over single pulses for two reasons: 1) to increase the total energy deposited during each perturbation, and 2) to ensure the formation of higher-temperature spark discharges as opposed to glow discharges, which can occur if single pulses are used \citep{mckiernan_instability_2020}. Different pulse durations, 20 and 40~ns, were used to vary the energy deposited per pulse. Sample voltage and current traces from the first four pulses within a packet, along with their averages, are shown in figure \ref{pulseIV}, recorded with the model in the wind tunnel during a test at at $Re_{\infty} = 6.4~\times 10^6$~/m.  The overall pulse shapes and characteristics are consistent across each of the sample pulses, with the 40~ns pulses reaching a higher maximum current and voltage relative to the 20~ns pulses. The voltage and current reflections observed for the 40~ns pulses are attributed to an impedance mismatch between the pulse generator and the electrodes.  

Table \ref{tab:table1} contains the packet parameters pertaining to the four perturbation cases tested. The cumulative pulse energies are calculated by integrating the instantaneous power output, which is the product of the average current and voltage traces shown in figure \ref{pulseIV}. It is important to note that not all of the pulse energy measured in the circuit is transferred to the internal energy of the gas; some energy is dispersed through other processes such as ionization and emission \citep{pai_nanosecond_2010}. The measured pulse durations for the nominally 20 and 40~ns pulses, as determined by the full duration at half maximum (FDHM) power, are 21.6~ns and 41.9~ns respectively. The average individual pulse energy for the 40~ns pulses (0.522~mJ) is more than double that of the 20~ns pulses (0.220~mJ). This difference is one way that the perturbation energy is varied within the test campaign. As the pulse repetition frequency is held constant at 100~kHz, varying the number of pulses within the packet also varies the packet duration. Cases 2 and 3 have similar packet energies, with Case 2 having a longer packet duration and less energy per individual pulse. The effects of Cases 2 and 3 are compared to determine whether perturbation duration or individual pulse energy is more influential over the SWBLI unsteadiness. The average power released during a test, calculated as the product of the packet energy and perturbation frequency, ranges from 0.55 to 26.1~W depending on the conditions. Respectively, these values represent 7.6~$\times10^{-5 }$\% and 3.6 $\times10^{-3 }$~\%  of the inviscid flow power, $\rho_{\infty}U_{\infty}^3A_{test-section}$ \citep{webb_control_2013}.


\begin{table}

\begin{center}
    \def~{\hphantom{0}}

\begin{tabular}{lcccccc}
Case&  Pulse width, ns\ & No. pulses & Total packet energy, mJ & Packet duration, $\upmu$s \\ \hline
1& 20& 5& 1.10& 50\\
2& 20& 10& 2.20&100\\
3& 40& 5& 2.61& 50\\
4& 40& 10& 5.22&100\\

\end{tabular}
\caption{Perturbation Packet Parameters }
\label{tab:table1}
\end{center}
\end{table}

\subsection{Schlieren-Based Tracking Algorithm} \label{sec:schler}
\begin{figure}[h]
 \centering
  \subfloat[]{\includegraphics[width=0.75\textwidth]{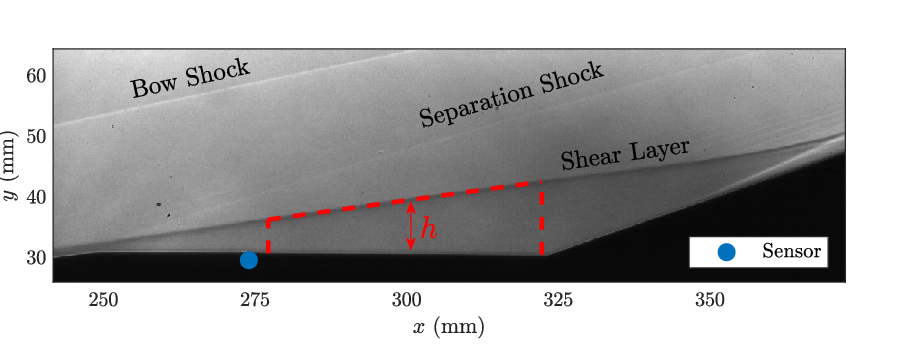}\label{schFOV:f1}}
  \hfill
  \subfloat[]{\includegraphics[width=.75\textwidth]{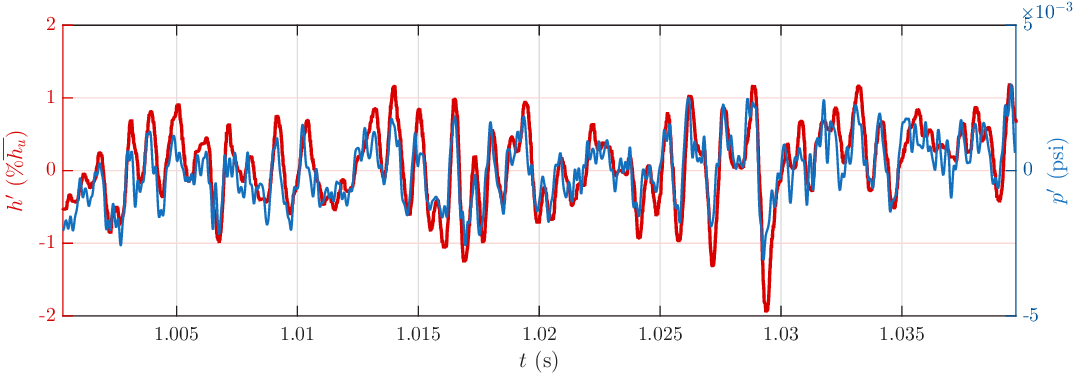}\label{schFOVg:f2}}
  
 \caption{a) Sample schlieren image showing the location of the Kulite sensor (blue dot) and the region used to determine separation height in the algorithm (red dashed lines). b) Comparison of separation height fluctuations tracked in the schlieren images with the low-frequency pressure fluctuations measured by the Kulite sensor inside of the separation bubble, for the case without plasma perturbations.}
 \label{SchFOV}
 \end{figure}

Despite mitigation efforts, sensor data were affected by EMI from the high-voltage pulse generator. Therefore, schlieren imaging is used as the main method to measure and analyze the effects of the perturbations on the separation bubble unsteadiness. An algorithm applied to the schlieren images identifies the edge of the shear layer through a weighted average of pixel intensity around the location of highest intensity for each column of pixels over the last 45~mm of the slice. The algorithm also identifies the ramp corner and slice locations in the schlieren images, using these as the basis of a coordinate system that ensures continuity across all analyses. A linear fit is applied to the shear layer coordinates, and the average distance between the slice and the shear layer is reported as the separation height for that frame. The separation height is tracked in this manner for at least 3000 consecutive schlieren frames per run, for a time period of approximately 40~ms. This is well within the 200-ms period of uniform flow conditions in the BAM6QT, while also allowing the low-frequency content of the separation unsteadiness to be well-resolved. For the lowest perturbation frequency tested, 500~Hz, this allows us to resolve 20 perturbation cycles. Because the separation front on this geometry does not move, the fluctuations in bubble height were chosen as a representative measurement of the bubble's unsteadiness, which oscillates as if it is hinged at the leading edge of the slice. All fluctuations in separation height are reported as the percentage difference from the time-averaged separation height in the unperturbed baseline case, $\overline{h_u'}$. Figure \ref{schFOV:f1} contains a sample schlieren image of the separation bubble, with the tracked region highlighted in red. Figure \ref{schFOVg:f2} compares the separation bubble fluctuations measured in the schlieren with the static pressure fluctuations measured by a Kulite sensor inside of the separation bubble during the baseline case. The data shown are the AC-coupled DC output from this sensor, which has been low-pass filtered to reveal only the low-frequency pressure fluctuations. The results from the schlieren-based separation-tracking algorithm agree very well with the pressure fluctuation data, with a decrease in separation height corresponding to a decrease in static pressure within the bubble. This confirms that the schlieren-based tracking algorithm is capable of accurately representing the separation unsteadiness when sensor data cannot be relied upon.

While 3D SWBLI cannot be characterized by a single length scale, the centerline separation length was chosen as a representative dimension in order to compare Strouhal numbers against existing literature. The calculation of Strouhal number based on a centerline length scale for 3D SWBLI has previously been applied to cylinder-induced separated flows \citep{combs_investigating_2018,ngoh_forced_2022}. Using the mean of 200 images, the reattachment location is approximated by extrapolating the visible shear layer to find its intersection point with the ramp. The centerline separation length, $L$, is defined as the axial distance between the separation point at the leading edge of the slice and the estimated mean reattachment point.

\section{Results}

\subsection{Baseline Flow}  
\begin{figure} [h!]
 \centering
  \includegraphics[width=0.85\linewidth]{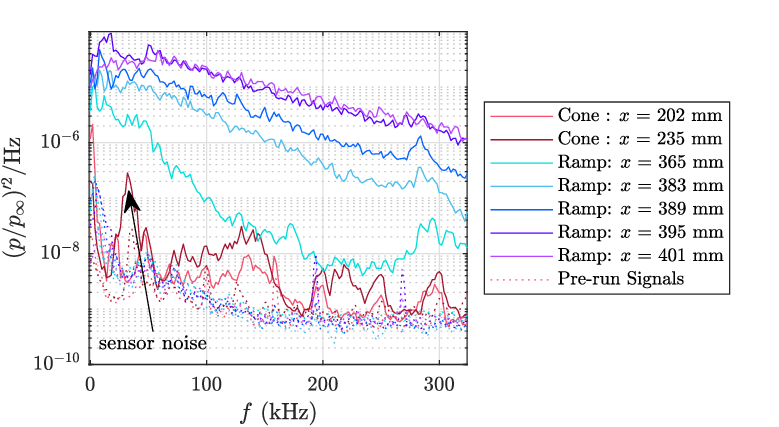}
  
 \caption{Power spectral densities of PCB pressure transducers on the cone upstream of the separation front and on the ramp around the reattachment region for the baseline flow [$Re_\infty = 6.35 \times 10 ^6 /m$]; see figure \ref{model} for sensor locations.}
 \label{PCB_PSD}
 \end{figure}

The unperturbed baseline flow behaviors are first characterized to provide a comparison against the flows forced by the NRP discharges. Figure \ref{PCB_PSD} shows power spectral density (PSD) analyses of pressure fluctuations measured on the surface of the cone and ramp. Fluctuations are measured by two PCB pressure transducers on the upper surface of the cone, which are located 202~mm and 234~mm downstream of the nose-tip, or approximately 42~mm and 10~mm upstream of the leading edge of the slice (see figure \ref{model}). The pre-run (no flow) signal is also analyzed to determine sensor noise levels. While the peak at 35~kHz is attributed to sensor noise, there is some energy content in the upstream boundary layer at frequencies up to 150~kHz, but the origin of this is not clear. Figure \ref{PCB_PSD} also includes the power spectra at five locations along the centerline of the ramp around the reattachment point, which has been estimated as $x\approx$ 380~mm for this configuration \citep{francis_effect_2024}. Therefore, the most upstream sensor location on the ramp is inside of the separated flow region. There is evidence of transition to turbulence in the reattached boundary layer, with broadband energy growing downstream of reattachment then remaining nearly constant between the two most downstream sensors.   

\begin{figure} [h!]
  \includegraphics[width=1\linewidth]{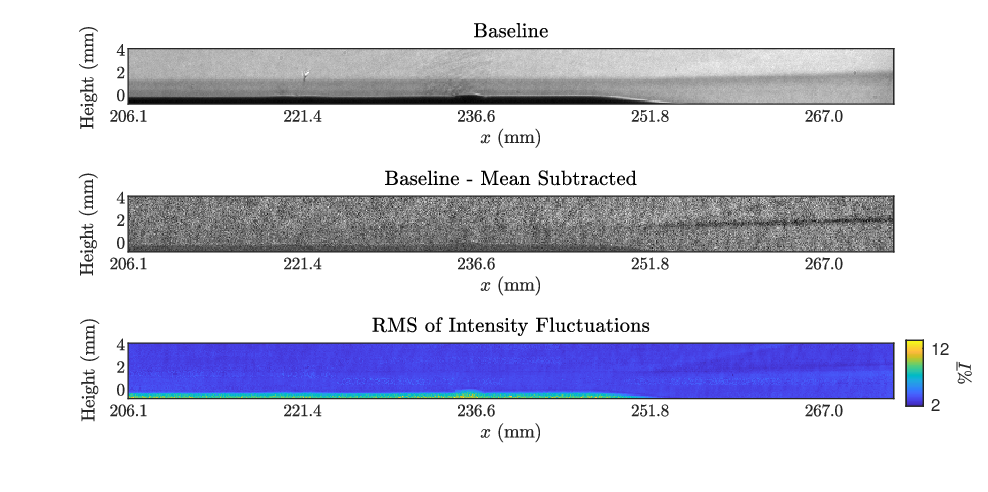}
   \centering
 \caption{Sample raw (top) and mean-subtracted (middle) schlieren images showing the state of the unperturbed boundary layer and upstream shear layer. The bottom image shows the root-mean-square (RMS) of the variations in pixel intensity as a percentage of the mean value for that pixel. The schlieren field of view is shown by the blue box in figure \ref{modelDim}.}
 \label{UnpSch}
 \end{figure}

The schlieren imaging of the baseline upstream boundary layer shown in figure \ref{UnpSch} exhibits no evidence of turbulent density fluctuations within the boundary layer. There are no visible structures in the unperturbed boundary layer, which is virtually indistinguishable from the flow outside of the boundary layer in the mean-subtracted image.  The wind tunnel model is visible in the mean-subtracted image because, during a run, the wind tunnel moves very slightly with respect to the fixed optical bench, so the model in the schlieren images is not completely still. Figure \ref{UnpSch} also contains an image of the RMS values for each pixel for a series of 1300 images at 875~kHz; there appears to be very little difference between the RMS values inside and outside of the boundary layer. The body of the wind tunnel model has higher RMS values, both because of the tunnel motion previously mentioned and because the pixel fluctuations are being represented as a percentage of the mean intensity value, which is much lower for the model region. The observations based on figure \ref{UnpSch} are indicative of a laminar upstream boundary layer for the baseline flow.

\begin{figure}[h!]
 \centering
  \subfloat[]{\includegraphics[width=.9\textwidth]{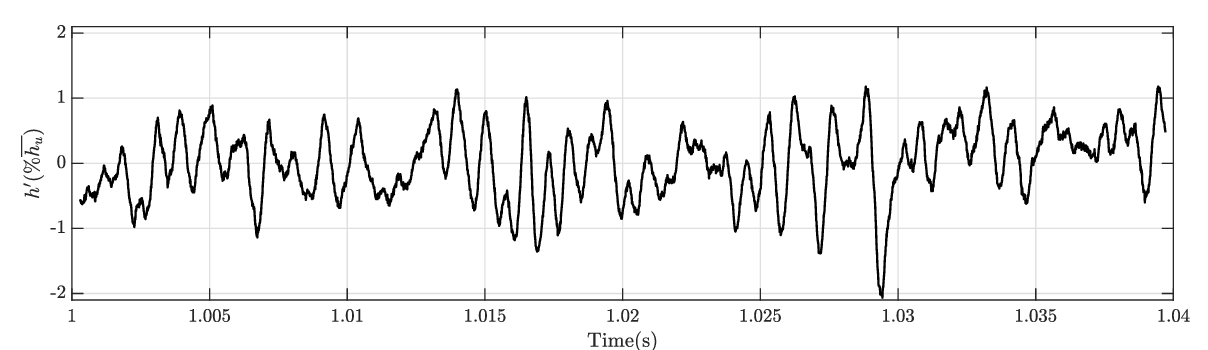}\label{Run15:f1}}
  \hfill
  \subfloat[]{\includegraphics[width=.45\textwidth]{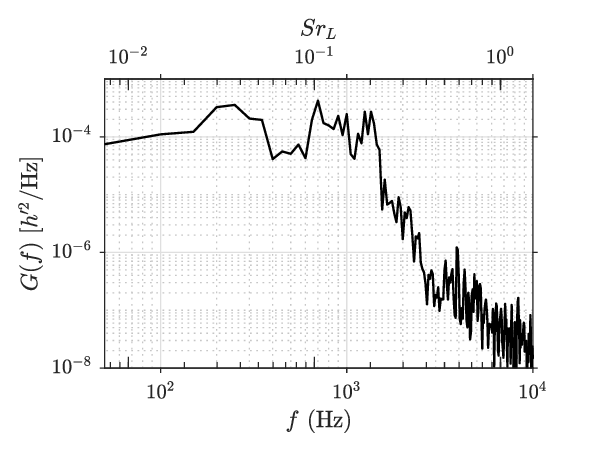}\label{Run15:f2}}
   \subfloat[]{\includegraphics[width=.45\textwidth]{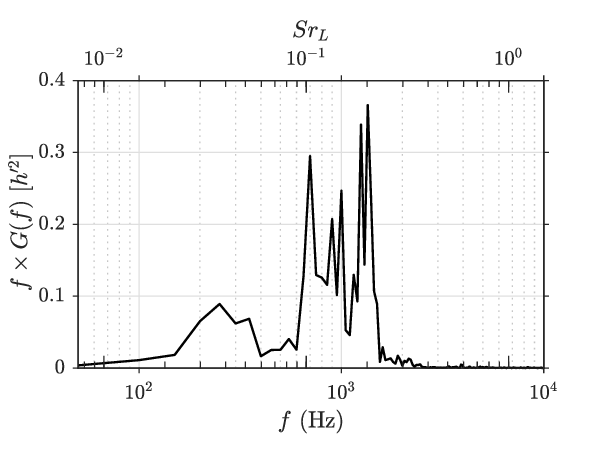}\label{Run15:f3}}
  
 \caption{(a) Separation-bubble height fluctuations derived from schlieren images of the baseline flow over a period of 40~ms. (b) Power spectral density analysis of the baseline separation height fluctuations, plotted with respect to both frequency and Strouhal number. (c) Frequency pre-multiplied power spectral density of the baseline separation height fluctuations.}
 \label{Run15}
 \end{figure}

The schlieren tracking algorithm is applied to a series of images of the unperturbed interaction region, and the natural oscillatory behavior of the separation bubble is observed. Figure \ref{Run15:f1} shows the separation height fluctuations over a 40~ms window, as a percentage of the mean separation height value for the unperturbed flow. For the lowest peak frequency of 250~Hz, this window spans ten oscillation cycles, with the 200-ms window of uniform flow conditions containing approximately 50 cycles. Although the fluctuations in the separation bubble are a small percentage of the average height, they still contain obvious low-frequency content. The power spectral density values of the fluctuations shown in figure \ref{Run15:f1} are plotted in figure \ref{Run15:f2}. These are calculated using Welch's method \citep{welch_use_1967} with a frequency resolution of 50~Hz and a 50\% window overlap. There greatest power is associated with frequencies below 1500~Hz, with a sharp decrease in frequency content outside of that range. The sharp decline in power content above 1500~Hz indicates that it may be a cutoff frequency for the system, according to the models theorized by  \cite{plotkin_shock_1975} and  \cite{touber_low-order_2011}. In the pre-multiplied spectra, the highest energy is associated with $Sr_L \approx 0.1 - 0.2$ ($f \approx 700 - 1300$~Hz), which is within the mid-frequency band corresponding to secondary bubble oscillations in 2D turbulent interactions \citep{gaitonde_dynamics_2023}. Studies by \cite{adler_unsteadiness_2018} of turbulent 3D SWBLI with open flow separation found that unsteadiness in the mid-frequency band was more dominant than the low-frequency band; they related this mid-frequency unsteadiness to Kelvin-Helmholtz shedding from the separated shear layer. Another smaller peak occurs at $Sr_L=0.038$ which is within the range of Strouhal numbers commonly associated with large-scale bubble breathing for both transitional and turbulent SWBLI ($Sr_L$ = 0.03 -- 0.04) \citep{sansica_instability_2016,combs_investigating_2018,threadgill_transitional_2021,gaitonde_dynamics_2023}. Upon observing these Strouhal numbers in transitional flows, both \cite{sansica_instability_2016} and \cite{threadgill_transitional_2021} remarked that it implies that the mechanisms driving the unsteadiness are independent of incoming boundary-layer state. Furthermore, unsteadiness at this Strouhal number on the cone-slice-ramp is notable because the low-frequency unsteadiness for the flow over this geometry is a large-scale flapping motion, rather than a breathing motion. This implies that the low-frequency unsteadiness at $Sr_L=$ 0.03--0.04 is not exclusive to the bubble breathing motion. One explanation is that the flapping originates from the same mechanism that causes the breathing motion in other fully-separated flows, but the range of motion of the separation bubble is limited by the fixed separation point, resulting in bubble fluctuations in the form of flapping rather than breathing.

\subsection{Perturbed Flow}
\subsubsection{Spark-Discharge-Generated Disturbances}
\begin{figure}[h!]
 \includegraphics[width=1\linewidth]{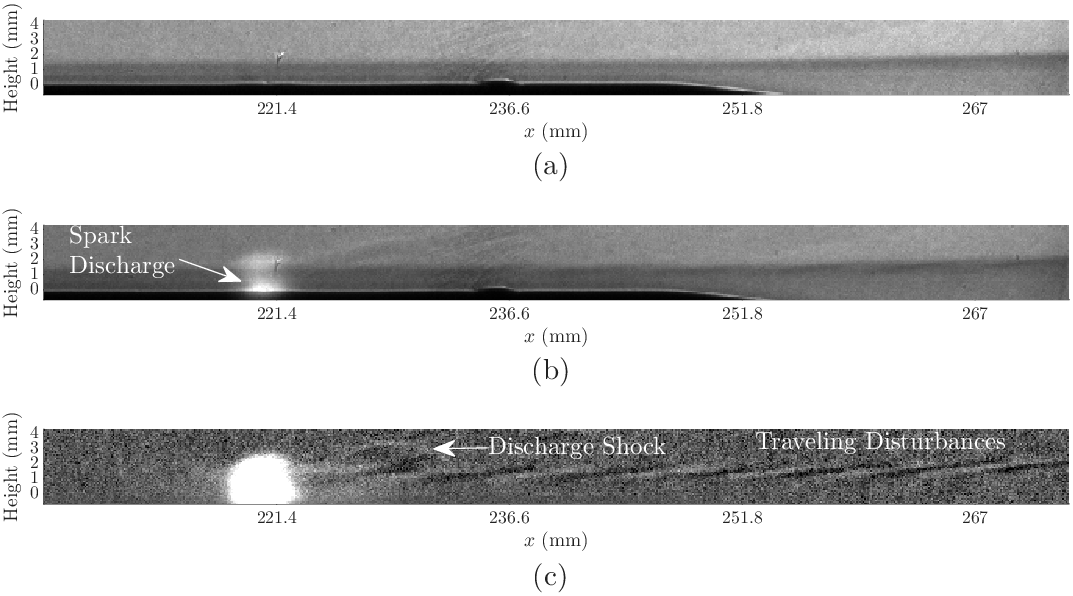}
 \centering
 \caption{Schlieren images of the region between the actuator and the start of the slice for (a) the baseline boundary layer; (b) the perturbed boundary layer 90~$\upmu$s after the first pulse in a packet (Case 4); and (c) the perturbed boundary layer 90~$\upmu$s after the first pulse in a packet (Case 4) [mean subtracted]. }
 \label{SchPlasma}
 \end{figure}

Packets of NRP plasma filaments are introduced into the boundary layer upstream of the SWBLI  at a range of perturbation frequencies and with varying energy input as discussed in Section \ref{sec:plasma_gen}. The perturbations begin 1~s into each test and last for 0.25~s. A single perturbation frequency is used for each test, with its packet conditions described by one of the four cases given in table \ref{tab:table1}. Many of the results presented here are from Case 4, which has the highest packet energy and therefore has the most pronounced effect on the flow. The effect of perturbations on the boundary layer downstream of the electrodes can be observed in the schlieren images in figure \ref{SchPlasma}. The field of view of the images includes the location of the electrodes at $x = 221$~mm, as well as the slice leading edge at $x = 248$~mm. Figure \ref{SchPlasma}a shows the baseline boundary layer which separates around $x = 250$~mm, resulting in the formation of a shear layer over the separated flow region. Figures \ref{SchPlasma}b and \ref{SchPlasma}c show schlieren images of the perturbed boundary layer without and with mean-subtraction, respectively. The perturbation packet shown in these images is described by Case 4, so the packet contains 10 pulses with pulse duration of 40~ns. The images in figures \ref{SchPlasma}b and  \ref{SchPlasma}c are from 90~$\upmu$s after the first pulse and thus show the formation of the tenth and final spark discharge in the packet. The mean-subtracted image of the boundary layer with active perturbation shows a series of disturbances convecting downstream through the boundary layer, as well as a weak shock caused by the most recent discharge. The boundary layer upstream of the actuator contains no visible fluctuations, which is consistent with the unperturbed boundary layer shown in figure \ref{UnpSch}. In figures \ref{SchPlasma}b and  \ref{SchPlasma}c, the disturbances can be seen traveling through the separation point and along the shear layer above the separation bubble. There is no visible change in separation location, which remains pinned at the leading edge of the slice. Although studies from \cite{sansica_instability_2016} and \cite{priebe_low-frequency_2012} suggest that the the laminar upstream boundary layer and large separation scale already diminish the influence of the upstream mechanism, the fixed separation point reduces it further, as the upstream disturbances cannot excite unsteady motion by convecting the separation point. 


\begin{figure}[h!]
 \centering
 \subfloat[Baseline]{\includegraphics[scale=.75]{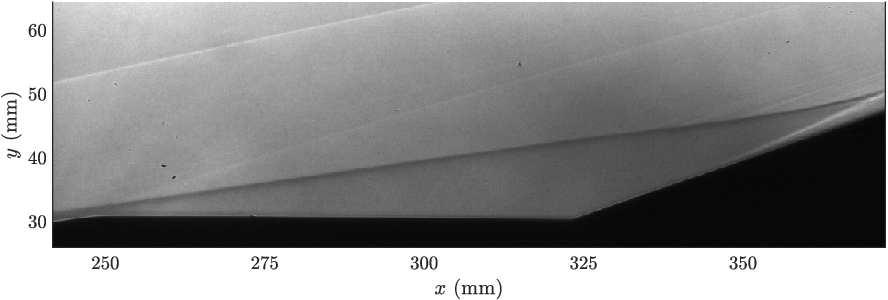}}
 
  \subfloat[Case 4 Perturbation Packet]{\includegraphics[scale=.75]{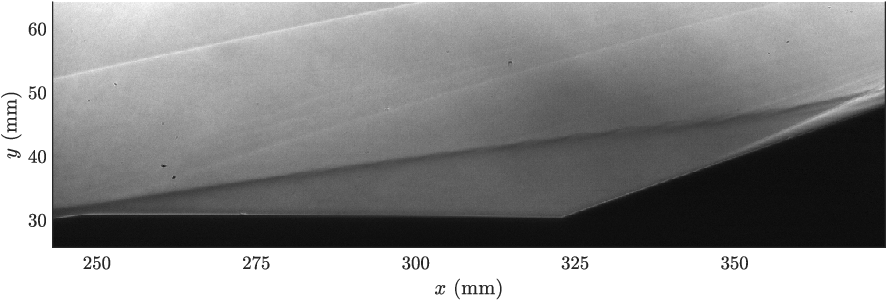}}
  
   \subfloat[Case 4 Perturbation Packet (mean-subtracted)]{\includegraphics[scale=.75]{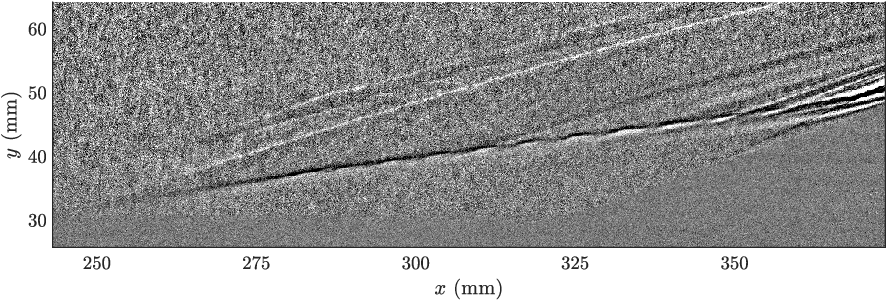}}
   
 \centering
 \caption{Schlieren images of the slice-ramp region showing the shear layer with and without traveling disturbance packets. Schlieren videos of the perturbation period are included as supplemental movies.}
 \label{SchPlasmaShearLayer}
 \end{figure}

Figure \ref{SchPlasmaShearLayer} contains schlieren images of the separation region with and without traveling perturbations. The image of the perturbed shear layer is from approximately 74~$\upmu$s after the last pulse in the packet was generated, and it shows the disturbances convecting through the shear layer over the separation bubble. The series of weak shocks generated by the formation of the spark discharges can be seen above the oblique shock, convecting downstream along with the disturbances. For each perturbation case and frequency, the separation-bubble height in the schlieren images is tracked according to the method described in Section \ref{sec:schler}. When the disturbances pass through the schlieren field of view, as seen in figure \ref{SchPlasmaShearLayer}, they disrupt the tracking algorithm and cause an artificially high value for the separation-bubble height in that frame. To address this, the height fluctuations for the perturbed cases are treated with a Hampel filter. This filter identifies and removes anomalies by comparing a datum with the median of its surrounding points; an outlier is defined using a threshold based on the median absolute deviation value. However, this does not always correct the issue entirely, and occasionally there is a slight bump or plateau in the signal during the time period when the plasma is passing through the field of view, which will be identified when relevant.

\begin{figure}
    \centering
    \subfloat[]{\includegraphics[width=0.4\linewidth]{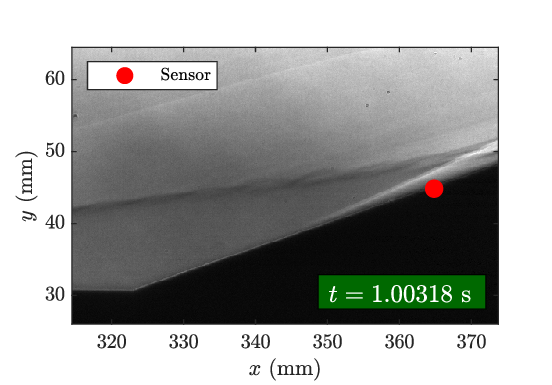}\label{dista}}
    \subfloat[]{\includegraphics[width=0.4\linewidth]{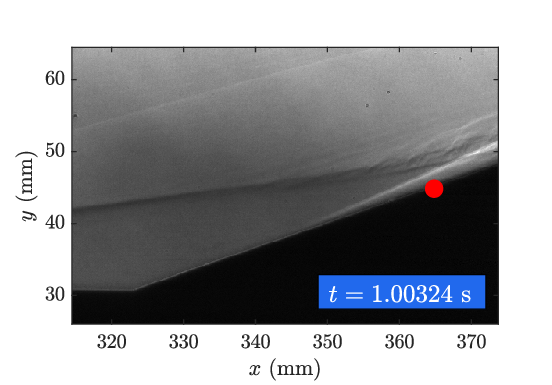}\label{distb}}
    
    \subfloat[]{\includegraphics[width=0.4\linewidth]{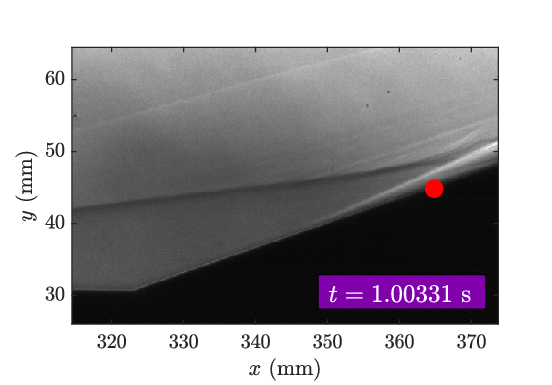}\label{distc}}

    \subfloat[]{\includegraphics[width=1\linewidth]{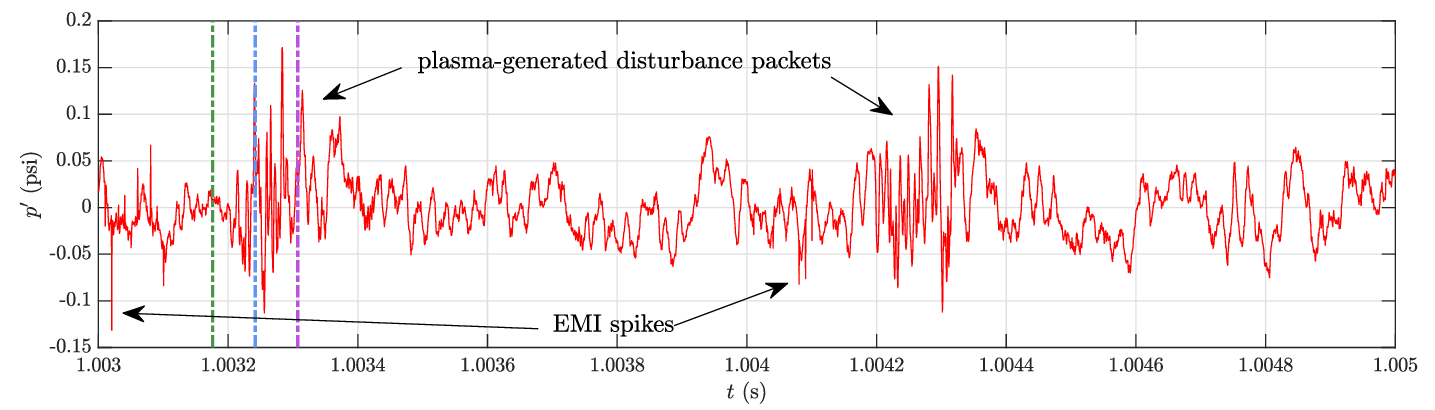}\label{press}}
    
    \subfloat[]{\includegraphics[width=1\linewidth]{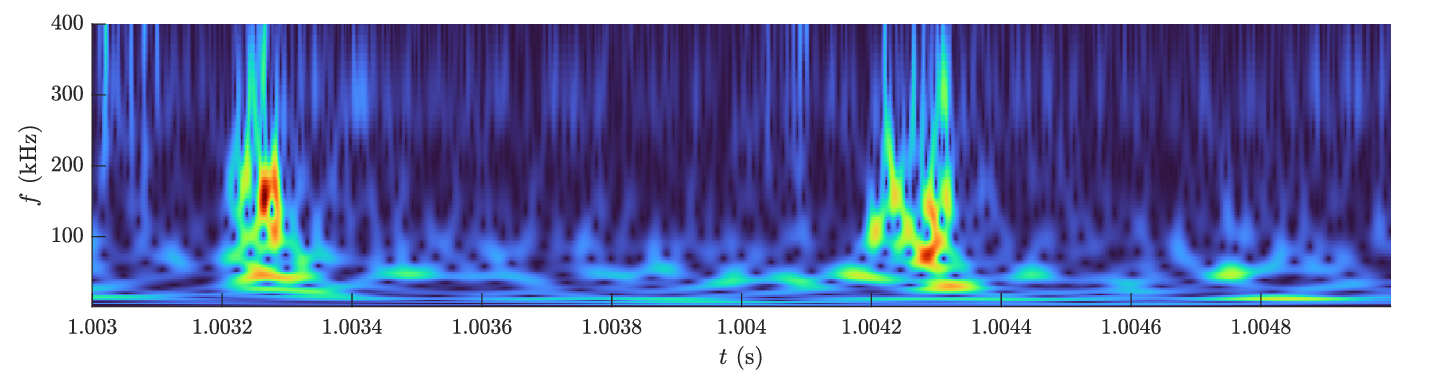}\label{CWT}}
    \caption{(a-c) Schlieren images showing the disturbance packet traveling through the shear-layer reattachment point at the ramp. The location of the pressure sensor on the ramp is indicated in red. (d) Pressure fluctuations measured on the ramp's surface for two perturbation cycles ($f_p = 1000$ $\mathrm {Hz}$, Case 4). The dashed lines indicate the times corresponding to the above images. (e) A continuous wavelet transform of the pressure fluctuation data shown in (d).}
    \label{fig:disturbances}
\end{figure}

Figure \ref{fig:disturbances} shows the behavior of the traveling perturbations as they pass through the reattachment region, causing periods of turbulent breakdown in the reattaching shear layer. The schlieren images in figures \ref{dista} - \ref{distc} show the reattaching shear layer at three times during the period when the traveling disturbances pass through reattachment. The red dot indicates the axial location of a PCB pressure transducer on the ramp; pressure fluctuations measured by this sensor throughout a period of two perturbation cycles are shown in figure \ref{press}. Sharp spikes which are identified as EMI from the high-voltage pulsing precede the arrival of the traveling disturbances, which introduce turbulent pressure fluctuations in the downstream shear layer. The continuous wavelet transform (CWT) of these pressure fluctuations is calculated using a Morlet waveform and shown in figure \ref{CWT}. The CWT shows that the perturbations cause periods of increased broadband turbulence without a consistent dominant frequency. The authors suggest that the mechanism by which these perturbations impact the SWBLI unsteadiness is related to the theory by \cite{priebe_low-frequency_2012} in which modulated turbulence levels in the shear layer affect the entrainment rate. They used this theory to describe the origins of unsteadiness in fully turbulent flows, but a similar mechanism could be at play for the perturbed flow. In the current work, the disturbances may be altering the entrainment rate by increasing vorticity \citep{tian_mechanisms_2021} as they pass through the shear layer and prompting turbulent breakdown around reattachment; this would lead to a cycle of mass entrainment and recharge that initiates unsteady motion in the separation region. 


\subsubsection{Perturbed SWBLI Unsteadiness}

\begin{figure}[h!]
 \includegraphics[width=1\linewidth]{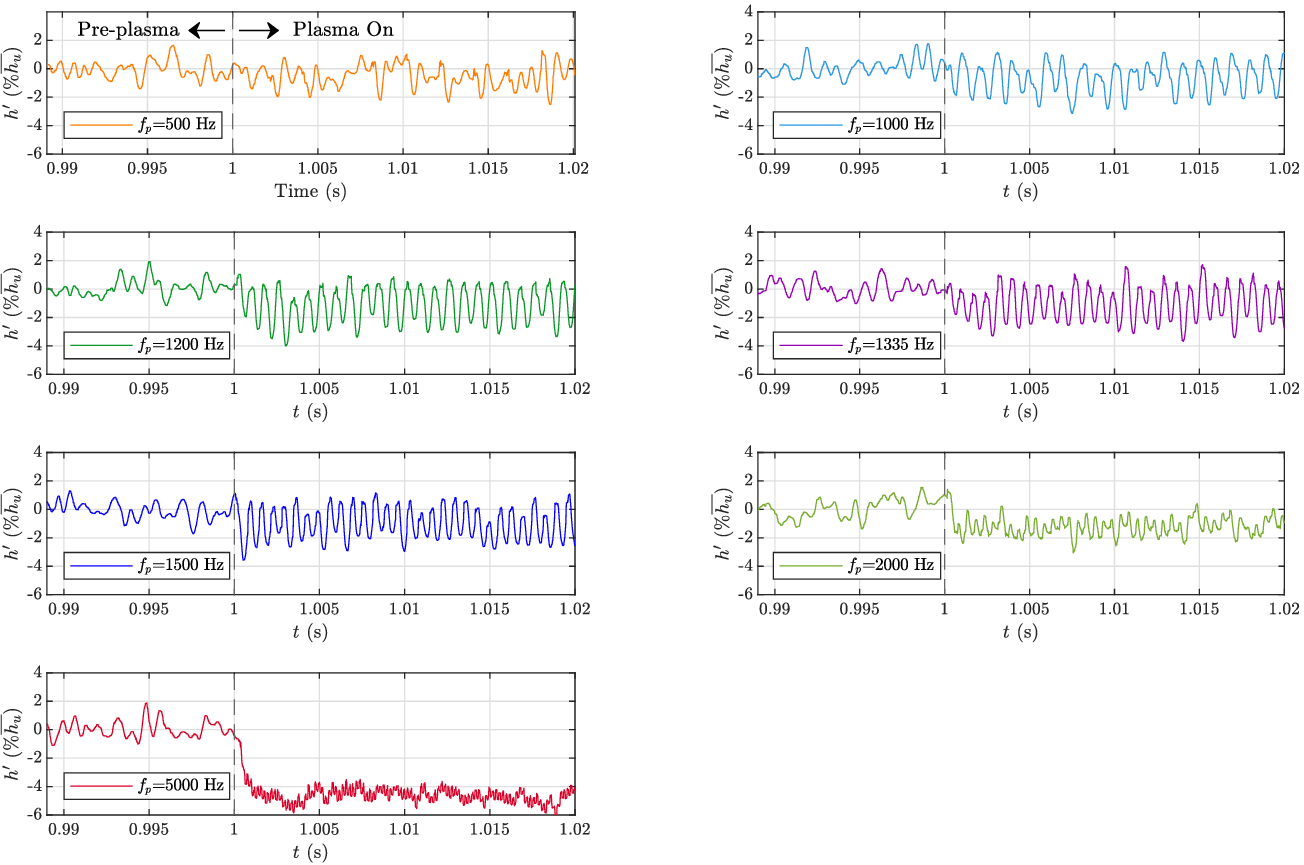}
 \centering
 \caption{Fluctuations in separation-bubble height before and after the start of the plasma perturbation period, as a percentage of the mean baseline height. Perturbation frequencies are between 500 and 5000~Hz, and packet conditions are described by Case 4. The introduction of the spark discharges begins at $t$ = 1~s.}
 \label{pre and post}
 \end{figure}

In each run, the SWBLI system is allowed to oscillate naturally until $t=$ 1 s, at which point the plasma perturbations begin at the selected frequency and packet conditions for that test. Plots of the separation bubble fluctuations for each perturbation frequency are shown in figure \ref{pre and post}. The time range shown on the plots contains both pre-plasma ($t<1$~s) and post-plasma ($t>1$~s) fluctuations. All results in figure \ref{pre and post} are from tests using the Case 4 packet conditions, which had the most prominent effects on the SWBLI unsteadiness. The dashed vertical line on each plot indicates the time when the plasma perturbations begin, after which the unsteady motion of the separation bubble is visibly altered. The 500~Hz perturbation frequency does not appear to induce oscillations at 500 Hz, but rather a higher frequency and there is no significant change in the mean height. For frequencies between 1000 and 1500~Hz, the separation bubble immediately responds to the perturbations, appearing to oscillate at the corresponding perturbation frequency with a slight decrease in the average separation height. When perturbations are introduced at 2000~Hz, there is an obvious decrease in the average height of the bubble, and the amplitudes of the oscillations are smaller than those at the lower perturbation frequencies. The 5000~Hz perturbation frequency causes the most dramatic decrease in separation height, and the separation unsteadiness exhibits lower-amplitude, higher-frequency oscillations compared to its natural behavior. This result is similar to what \cite{narayanaswamy_control_2012} observed when perturbing a turbulent SWBLI using pulsed plasma jets in the upstream boundary layer. When perturbing the flow at frequencies higher than the natural unsteadiness, they observed reduced oscillation amplitudes and a downstream shift in the mean shock position. 

\begin{figure}[h!]
 \includegraphics[width=1\linewidth]{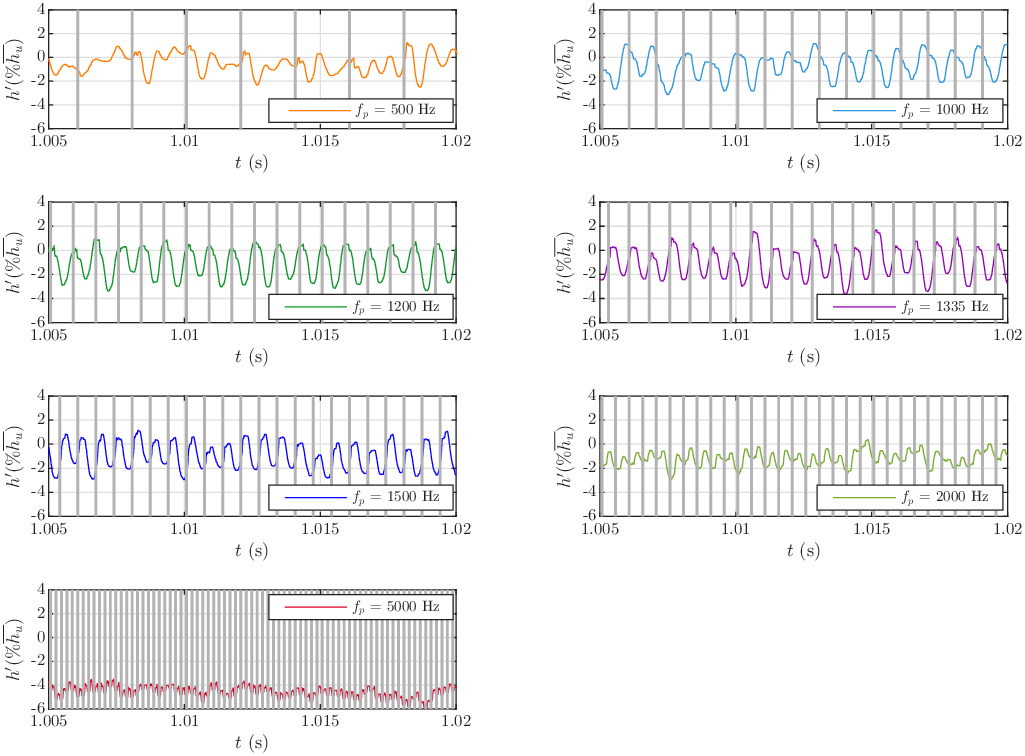}
 \centering
 \caption{Separation-bubble height oscillations during active plasma pulsing for a range of perturbation frequencies, for the highest-energy packet conditions (Case 4). The gray blocks indicate the time periods when the disturbances are passing through the shock foot.}
 \label{MeanDistances}
 \end{figure}

Plots of the separation height fluctuations for 15~ms within the plasma actuation period are shown in figure \ref{MeanDistances}. The grayed out regions indicate the time periods when the perturbations are passing through the SWBLI. The aforementioned issues caused by the shear layer disturbances in the schlieren field of view appear as slight bumps or plateaus in the separation heights immediately after the grayed out regions. For perturbation frequencies above 500~Hz, the separation bubble unsteadiness appears locked to the frequency that the plasma packets are introduced into the upstream boundary layer. There is an obvious decrease in oscillation amplitude and average separation height at $f_p= $~2000~and~5000~Hz. Notably, these frequencies are above the apparent cutoff frequency of the separation bubble's natural oscillations, as seen in figure \ref{Run15}. As perturbation frequency increases, the duty cycle of the perturbations increases, up to $f_p = 5000 $~Hz, at which point the packets are present 50\% of the time. The behaviors seen in figure \ref{MeanDistances} persist throughout the entire 0.25~s period of plasma perturbation. 

\begin{figure}[h!]
  \subfloat[]{\centering\includegraphics[width=0.5\linewidth]{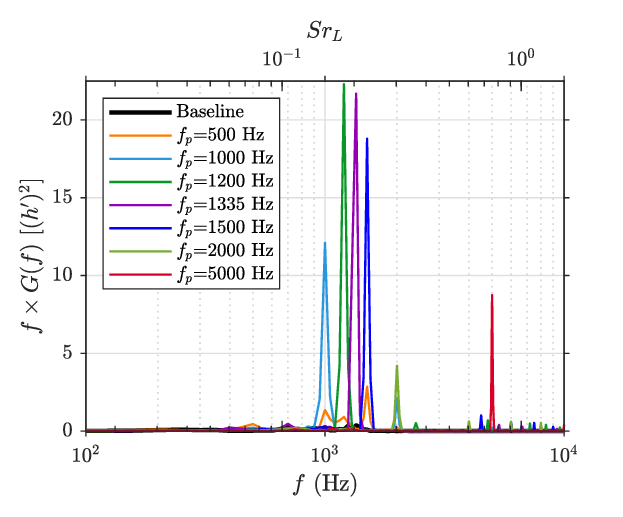}\label{psd4:f1}}
  \hfill
  \subfloat[]{\centering\includegraphics[width=0.5\linewidth]{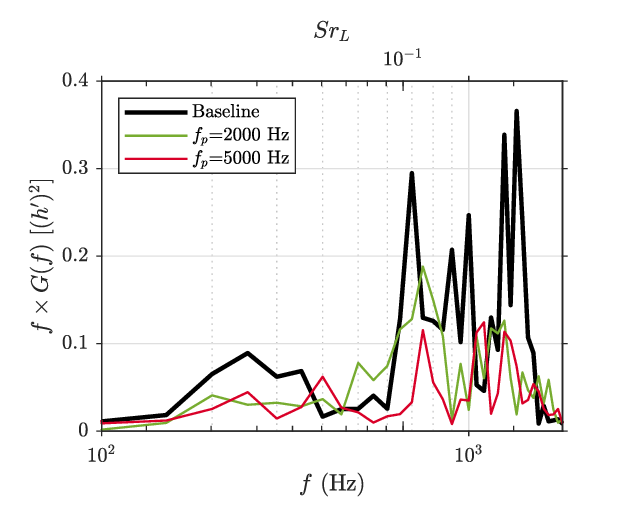}\label{psd4:f2}}
  
 \caption{Power spectral densities of the separation-bubble fluctuations in baseline and perturbed flow (Case 4) with (a) all perturbation frequencies, and (b) a detailed view of the low-frequency effects of the two highest perturbation frequencies. }
 \label{PSDcase4}
 \end{figure}

Figure \ref{psd4:f1} shows the pre-multiplied power spectra of the separation fluctuations in figure \ref{MeanDistances} compared with the unperturbed baseline case. The power spectra for perturbation frequencies between 1000 and 1500~Hz exhibit significantly higher energies at their respective frequencies than the power spectra for perturbation frequencies outside of that range. It is clear that perturbation within this frequency range is eliciting a much stronger response than perturbation outside this range, with $f_p$ = 1200 and 1335~Hz having the two highest peaks in power. Figure \ref{psd4:f2} contains a detailed view of the low-frequency effects of the 2000~Hz and 5000~Hz perturbations, compared with the baseline case. The higher-frequency perturbations cause a decrease in the power associated with almost all frequencies below 1500~Hz, with noticeable reductions in the baseline case's most prominent peaks at 250, 700, 1250, and 1350~Hz. In these cases where the perturbation frequency is above the cutoff frequency of the baseline oscillation envelope, the energy at those natural oscillation frequencies is reduced, with the effect becoming more pronounced as $f_p$ increases. This effect is not observed for perturbation frequencies less than or equal to 1500~Hz.   

\begin{figure}[h]
    \centering
  
    \includegraphics[width=1\linewidth]{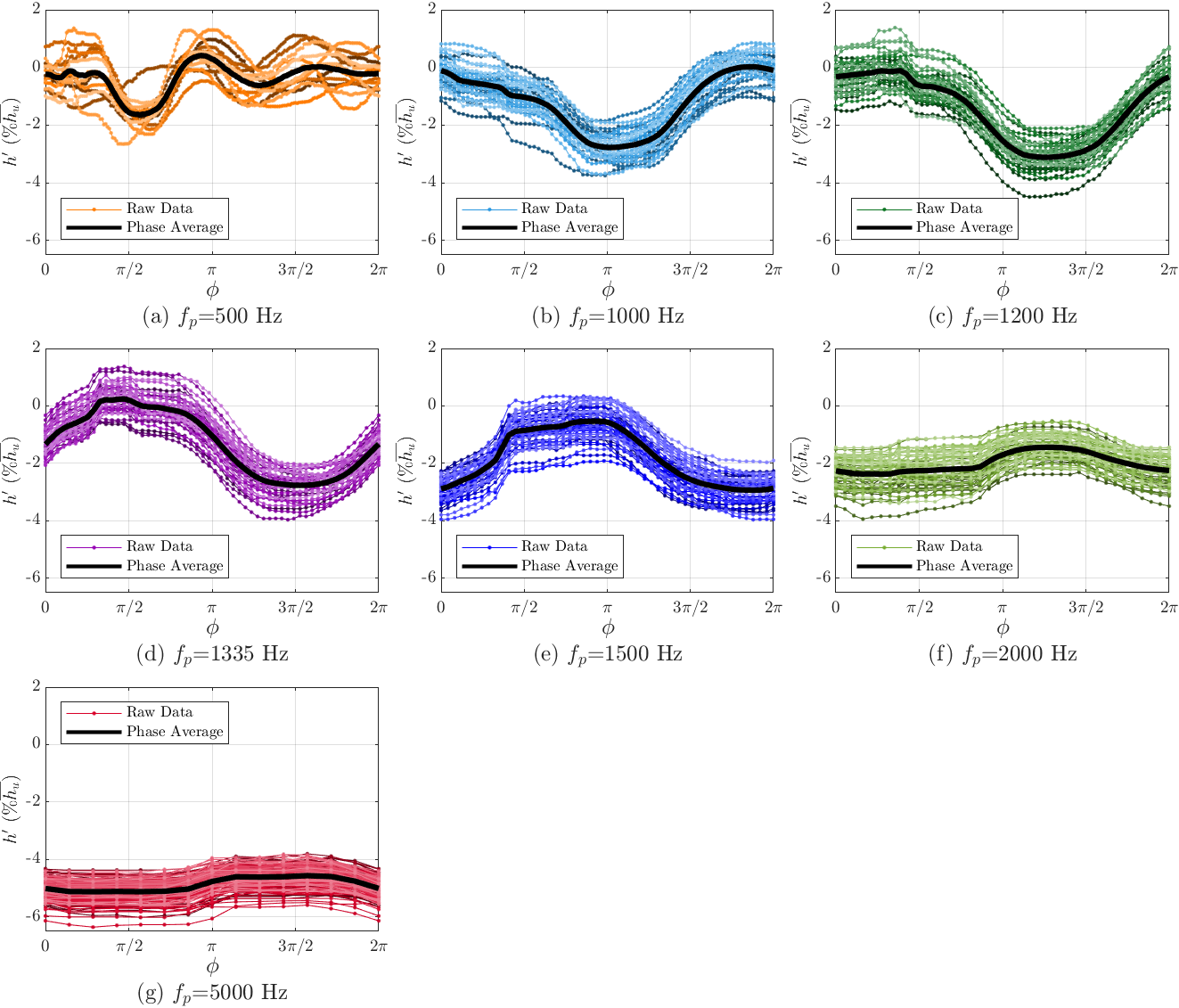}
    \caption{Phase averages of separation-height fluctuations for Case 4, as a percentage of the mean unperturbed separation bubble height, for perturbation frequencies between 500 and 5000~Hz.}
    \label{PhaseAvgs4}
\end{figure}

To visualize the mean effects of plasma perturbations, the separation height fluctuations are phase-averaged according to their respective perturbation frequencies. The traces of the height fluctuations, like those shown in figure \ref{MeanDistances}, are split into fragments the length of one period of the relevant perturbation frequency. The beginning of the fragment, corresponding to the phase $\phi$ = 0, is the time at which the first plasma pulse in a packet reaches the SWBLI region. The precise time is determined using the convection time obtained from the high speed schlieren visualization, along with the known pulsing times for each test. The end of a fragment, or $\phi = 2\pi$, corresponds to the time when the first plasma pulse in the subsequent packet reaches the SWBLI. Since the perturbation frequencies were not always a factor of the schlieren frame rate, a linear interpolation is used so that the fragment starts and ends at the exact time of the packets' arrival. Overlaying all of these fragments and taking their average at each phase reveals any behavior in the signal at the phase average frequency. The phase averages for all perturbation frequencies with Case 4 packet conditions are plotted in figure \ref{PhaseAvgs4}, alongside the raw data that make up the individual fragments. The fluctuation data are not centered about zero because they are presented as fluctuations from the mean baseline separation height, and the mean height of the bubble varies with separation frequency, as seen in figure \ref{pre and post}. The phase averages show that the separation unsteadiness responds differently depending on the perturbation frequency, with amplitude, mean height, and waveform shape all varying. The artifacts resulting from the disturbances in the schlieren frame appear in the phase averages as slight raised regions, such as the one seen just before $\phi = \pi/2$ in figure \ref{PhaseAvgs4}d; these do not represent any actual flapping in the separation bubble.

\begin{figure}[h!]
 \includegraphics[width=1\linewidth]{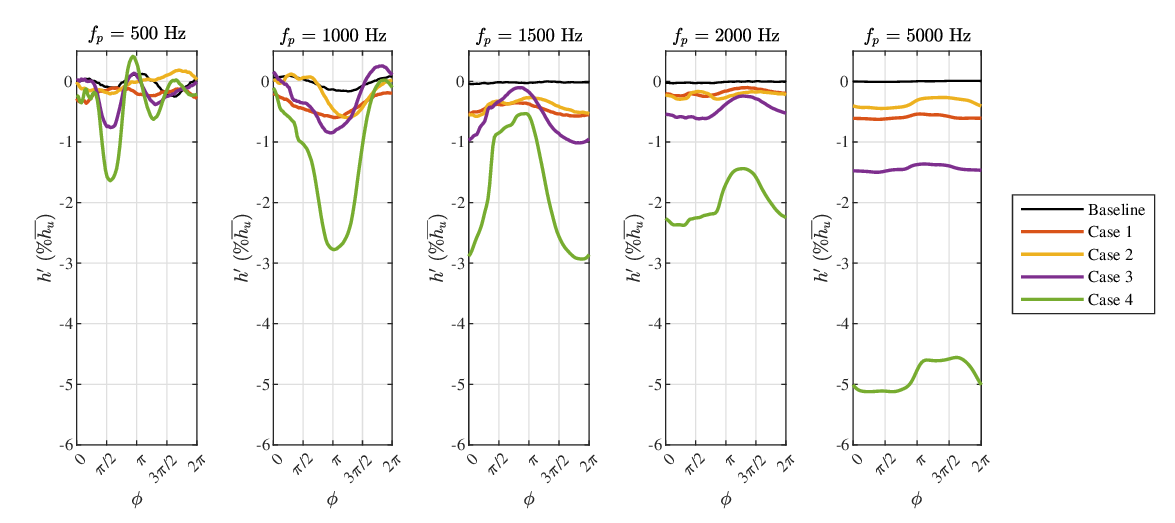}
 \centering
 \caption{Phase averages of separation height fluctuations for all plasma packet cases at each perturbation frequency.}
 \label{PhaseAllCase}
 \end{figure}

The phase-averaging process is repeated for the measured separation height fluctuations for all plasma packet cases, and the results for each perturbation frequency are shown in figure \ref{PhaseAllCase}. The fluctuations of the unperturbed separation bubble are also phase averaged at each perturbation frequency and are included as a baseline for comparison. In general, the shapes of the phase-averaged curves are similar for each case within a single perturbation frequency, but their amplitudes and offsets vary significantly depending on the plasma packet case being tested. For all perturbation frequencies besides 500~Hz, the phase averaging reveals that the mean effect introduced by the plasma is a quasi-sinusoidal oscillation at the perturbation frequency. For $f_p$= 500~Hz, Cases 1 and 2 do not have a clear impact on the separation oscillation, and their phase averages are shaped differently from one another and from those of Cases 3 and 4. However, the phase averages at $f_p$= 500~Hz for the two highest-energy packet cases contain multiple oscillations which decrease in amplitude.
 
\begin{figure}[h!]
 \includegraphics[width=0.65\linewidth]{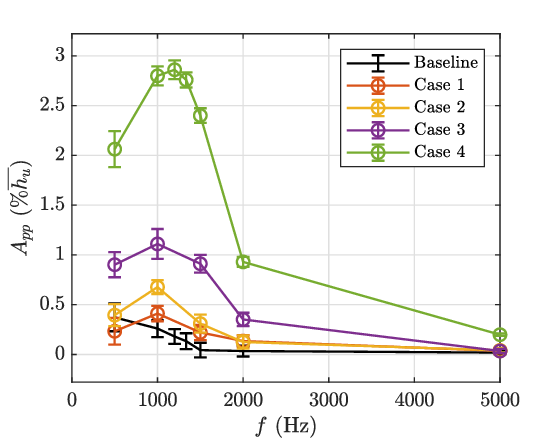}
 \centering
 \caption{Effect of perturbation frequency on phase average amplitudes for each plasma packet case.}
 \label{PhaseAmp}
 \end{figure}

  It is clear in figure \ref{PhaseAllCase} that the amplitudes of the phase-averaged fluctuations vary significantly with both frequency and case number. The phase-averaged amplitude values, $A_{pp}$, are determined by taking the difference between the maximum and minimum within a phase average, with the following exception: the regions where the plasma disturbances cause an artificially high separation height measurement are not considered, except when $f_p$ = 5000~Hz, in which case it is impossible to avoid. As a result, it is possible that the real amplitudes for the 5000~Hz perturbations are even lower than what has been shown here. The peak-to-peak amplitudes of the phase averages are shown in figure \ref{PhaseAmp}. The relationship between perturbation frequency and amplitude is similar for all packet cases, with the highest amplitudes occurring at perturbation frequencies between 1000 and 1500~Hz, and a significant decrease in amplitude for $f_p  \geq 2000$~Hz. The amplitudes of the Case 4 phase averages for the 1200 and 1335~Hz perturbation frequencies are also plotted; similarly to the power spectral density results, these frequencies result in the two highest phase average amplitudes. It is evident that the separation unsteadiness responds more strongly to forcing at frequencies within its natural oscillation envelope. This behavior was observed in supersonic compression ramp computations by \cite{poggie_effect_2019}, where upstream forcing at the natural unsteadiness frequency yielded a significantly higher phase-average amplitude than forcing at ten times that frequency. The fact that the induced oscillation amplitudes are frequency-dependent supports the theory that the separation bubble has a preferred oscillation frequency \citep{clemens_low-frequency_2014} and/or a cutoff frequency \citep{plotkin_shock_1975,touber_low-order_2011}. This result is also consistent with the theory that the forced dynamics are tied to a weakly damped global mode \citep{porter_selective_2019, cura_linear_2025}. This mode has been studied through resolvent analyses of laminar 
\citep{bugeat_low-frequency_2022} and turbulent \citep{cura_low-frequency_2024} separation bubbles, where the relationship between frequency and optimal gain resembles that of a low-pass filter.

\subsubsection{Effect of Perturbation Packet Conditions}
\begin{figure}[h!]
 \centering\includegraphics[width=0.65\linewidth]{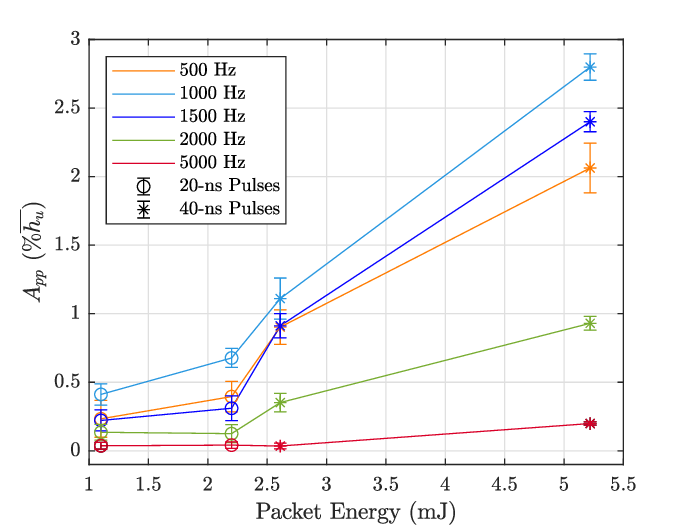}
   \caption{Effect of packet energy and perturbation frequency on peak-to-peak amplitude of the phase-averaged separation unsteadiness, as a percent of the mean unperturbed separation height.}
 \label{EnergyEffectsAmp}
 \end{figure}

Figure \ref{EnergyEffectsAmp} shows the relationship between the perturbation packet energy and the phase-average amplitudes shown in figure \ref{PhaseAmp}. While it is evident that higher packet energies result in greater phase-average amplitudes, it appears that the packet duration and individual pulse energy are also influential factors. Cases 2 and 3 have similar total packet energy (2.20 and 2.61 mJ, respectively) but Case 3 has half the packet duration and more than twice the individual pulse energy of Case 2. For all perturbation frequencies except for $f_p$ = 5000~Hz, the phase-average amplitudes for Case 3 were between 64\% and 190\% higher than for Case 2. Therefore, it appears that a stronger perturbation over a shorter period of time can excite the low-frequency unsteadiness more effectively than a weaker perturbation applied for a longer period of time. This is evidence that the duration of the disturbance is not necessarily the most important factor in exciting a low-frequency response. In fact, for all experiments, the time scales associated with the pulse durations, pulse frequency, and packet lengths are all one or more orders of magnitude below the time scales associated with the unsteadiness that they are able to excite.

\begin{figure}[h!]
  \subfloat[]{\centering\includegraphics[width=0.5\linewidth]{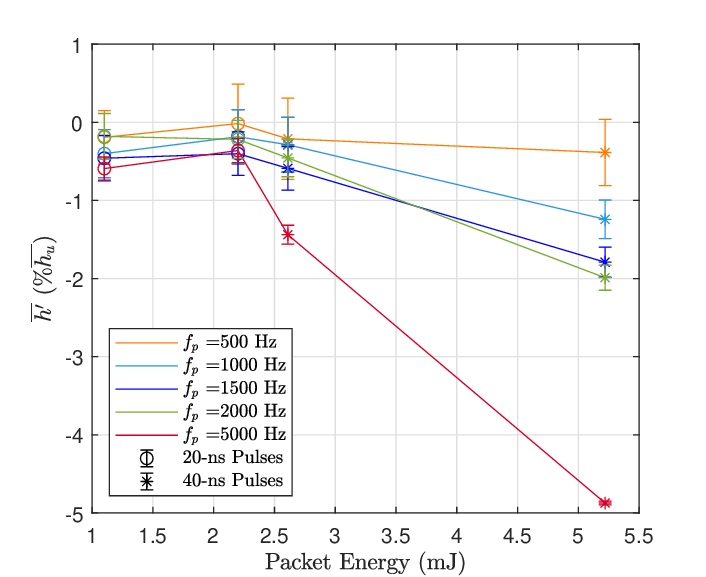}\label{EnergyEffects:f1}}
  \hfill
  \subfloat[]{\centering\includegraphics[width=0.5\linewidth]{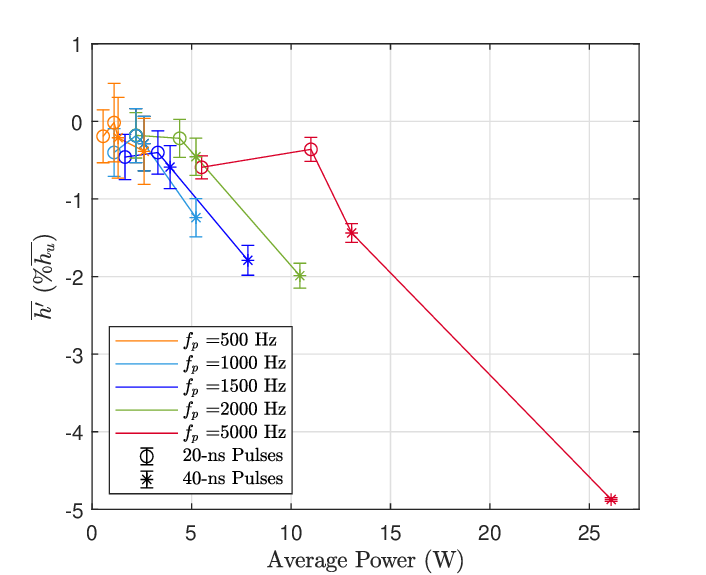}\label{EnergyEffects:f2}}
  
 \caption{Changes in mean separation bubble height caused by plasma perturbations, as the percent difference from the mean unperturbed separation height, with respect to (a) perturbation packet energy and (b) average power released during a test period.}
 \label{EnergyEffects}
 \end{figure}

From figures \ref{pre and post} and \ref{PhaseAllCase} it appears that the plasma perturbations can also affect the mean height of the separation bubble, with the magnitude of this effect varying with frequency and perturbation strength. Figure \ref{EnergyEffects} shows how the decrease in mean separation bubble height is affected by both perturbation packet energy and the time-averaged power released during the period of plasma actuation. In figure \ref{EnergyEffects:f1}, the mean separation height decreases as both packet energy and perturbation frequency increase. As mentioned previously, one theorized control mechanism of these high-temperature discharges is the introduction of vortices into the boundary layer. It is possible that the increased vorticity in the upstream boundary layer, and its entrainment of high-momentum fluid from the freestream, is responsible for the decrease in separation bubble size, with stronger and more frequent vortices resulting in a more significant decrease. Additionally, the disturbances may be increasing the mean entrainment rate in the shear layer and decreasing the mean bubble size by depleting mass through that mechanism. Figure \ref{EnergyEffects:f2} compares the change in separation height to the average plasma power during a test, calculated by taking the product of packet energy and perturbation frequency. Although higher average power generally yields a greater decrease in separation height, it is clear that there is also a dependence on the perturbation frequency. Perturbation at $f_p$ = 500~Hz yields the lowest average power values and causes no considerable decrease in separation height for all packet conditions. Also, the average heights are comparable for all points corresponding to Cases 1 and 2 (open circles), indicating that there could be a threshold packet energy required to affect the mean separation bubble characteristics. However, there is a clear negative linear trend between average power and separation-height change for Cases 2-4. Notably, these cases are offset from one another in increasing frequency order, with more power required by higher perturbation frequencies to achieve the same decrease in separation bubble height. 

\subsubsection{Perturbed Separation Bubble Dynamics}
\begin{figure}[h!]
 \includegraphics[width=0.65\linewidth]{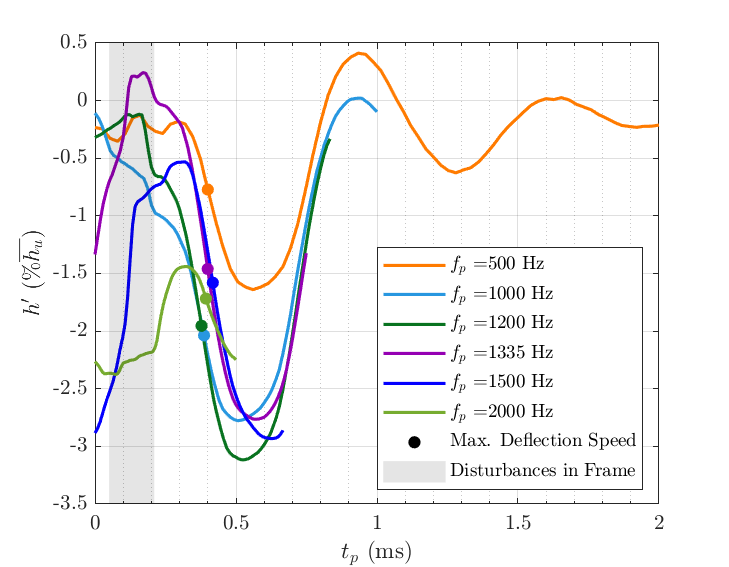}
 \centering
 \caption{The phase averages from Case 4 plotted with respect to time within the phase average period, $t_p$. Points where the maximum deflection speed occurs for each perturbation frequency are indicated by the solid circles in their respective colors. The grayed out region denotes the time period when the disturbances are visible in the schlieren field of view, resulting in a signal distortion not representative of actual separation motion.}
 \label{Response}
 \end{figure}

The phase averages from Case 4 are plotted on a common time scale in figure \ref{Response}, with  $\phi =0$ corresponding to $t_{p}= 0 $ and $\phi =2\pi$ corresponding to $t_{p}= 1/f_p $. Because the phase-averaging method factors in convection time, $t_{p}= 0 $ coincides with the time when the first disturbance in the perturbation packet reaches the SWBLI. The phase averages are analyzed on a common time scale in order to identify similarities in the effects of the perturbation packets at various frequencies. The grayed out area indicates the time period when the disturbances are traveling through the schlieren frame, affecting the readings from the schlieren-based tracking algorithm. After about 0.28~ms, the disturbances have traveled past the separation bubble entirely. In each of the phase averages, a downward deflection in the shear layer occurs about 0.4~ms after the start of the packet's passage through the SWBLI; $f_p$ = 5000~Hz is not included in this figure as its period is not long enough to compare with the behavior seen in the other phase-averages. Because this downward deflection occurs at roughly the same time in each of the phase averages, it appears that perturbation frequency does not influence the delay time of the separation response, which is herein defined as the time between the arrival of the first pulse at the SWBLI and when the maximum deflection speed occurs, indicated by colored circles on in figure \ref{Response}. The response delay times for all perturbation frequencies are between 0.378 and 0.418~ms. The periods of turbulent breakdown caused by disturbance packets in the reattaching shear layer occur when $t_p\approx 0.2-0.3$~ms (see figure \ref{press}), which is coincides with the time period in figure \ref{Response} when the separation heights first begin to decrease. It reasons that the delay time, which is similar for all perturbation frequencies, is tied to the convection time of the pulses, which does not very with perturbation frequency.
This suggests that the perturbations are activating a downstream mechanism, rather than an upstream one, consistent with the idea that the traveling disturbances are exciting the mass entrainment cycle.

\begin{figure}[h!]
\centering

\subfloat[]{\centering\includegraphics[width=.49\linewidth]{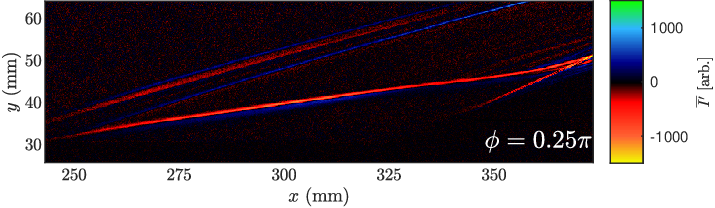}\label{25a}}
\hfill
\subfloat[]{\centering\includegraphics[width=.49\linewidth]{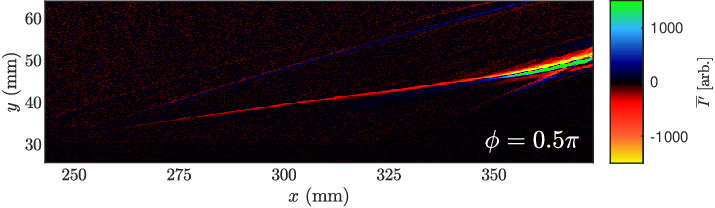}\label{25b}}

\subfloat[]{\centering\includegraphics[width=.49\linewidth]{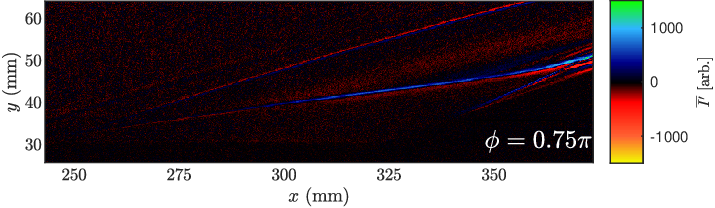}\label{25c}}
\hfill
\subfloat[]{\centering\includegraphics[width=.49\linewidth]{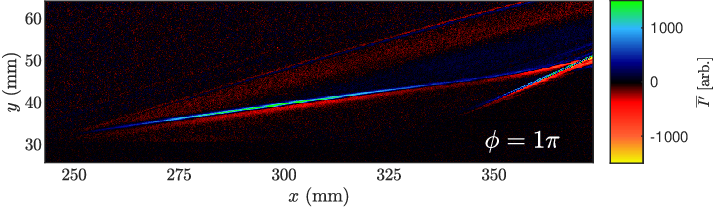}\label{25d}}

\subfloat[]{\centering\includegraphics[width=.49\linewidth]{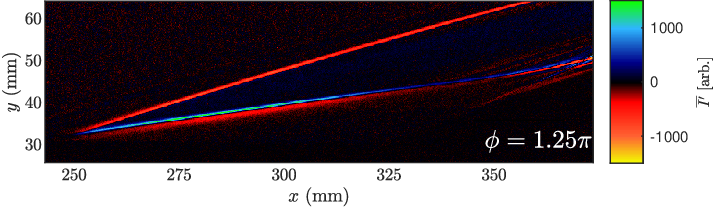}\label{25e}}
\hfill
\subfloat[]{\centering\includegraphics[width=.49\linewidth]{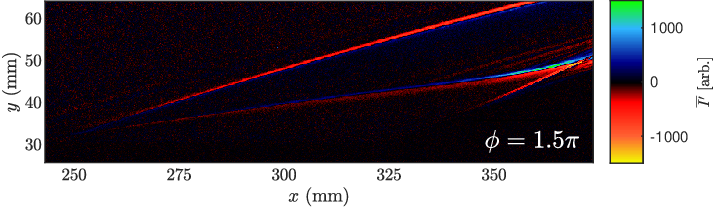}\label{25f}}


\subfloat[]{\centering\includegraphics[scale=.75]{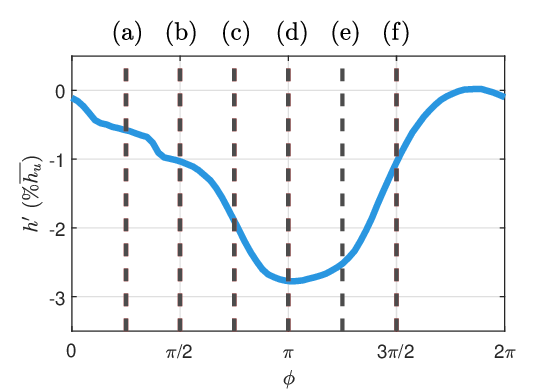}\label{25g}}
 \caption{Mean-subtracted phase-averaged schlieren images showing intensity fluctuations at different phases throughout the 1000-Hz perturbation case. The separation-height phase average is included in (g) for reference, along with dashed lines indicating the corresponding phase for each image shown.}
 \label{PhAvSch_1}
 \end{figure}

In order to further investigate the delayed reaction of the separation bubble shown in figure \ref{Response}, schlieren images are phase averaged using the same method applied to the height fluctuations, over the same time period. Phase-averaged intensity fluctuations are determined by subtracting the mean image of the entire set from the phase-average. Figure \ref{PhAvSch_1} shows intensity fluctuations at various phases throughout the $f_p$= 1000~Hz perturbation cycle. The corresponding phase average of the separation bubble height fluctuation is included in figure \ref{25g}, with vertical dashed lines each indicating the phase of the corresponding image in figures \ref{25a}-\ref{25f}. Although only certain phases are shown in this figure, a video showing the entire set is included in the supplemental material attached to this work. Figure \ref{25a} corresponds to the phase where the disturbance packet is traveling through the schlieren field of view. Because the shear layer appears as a dark region in the schlieren, the negative (red) line indicates that the shear layer at this phase is higher than its average value during this time period, which is consistent with figure \ref{25g}. The faint pair of red and blue streaks above the shear layer indicate the path of the spark-discharge-generated shocks as they convect downstream. Figure \ref{25b} shows the phase when the traveling disturbance packets are interacting with the reattachment region, with the increased turbulence seen in figure \ref{fig:disturbances} causing significant intensity fluctuations (yellow/green) where the shear layer meets the ramp. Figures \ref{25c} through \ref{25e} show that the decrease in shear layer height begins downstream ($\phi$ = 0.75 $\pi$) and travels upstream ($\phi$ = 1 $\pi$), with the separation shock motion following the separation motion ($\phi$ = 1.25 $\pi$). Because the separation shock appears as a light region in the schlieren, the decrease in intensity (red) in the shock region with the slight increase (blue) underneath indicates the shock moving downward ($\phi = 1.25-1.5 \pi$). From these images, it is clear that the turbulent breakdown near reattachment precedes the separation bubble motion which precedes the shock motion.

\begin{figure}[h!]
\centering
\subfloat[]{\centering\includegraphics[width=.49\linewidth]{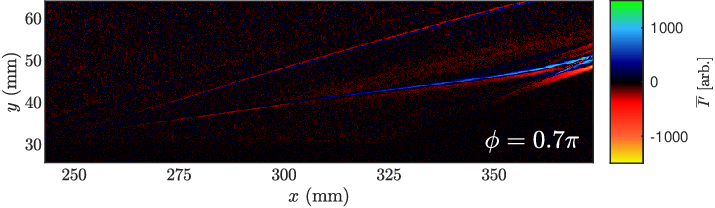}\label{25a2}}
\hfill
\subfloat[]{\centering\includegraphics[width=.49\linewidth]{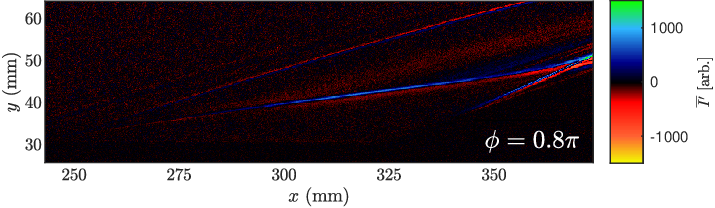}\label{25b2}}

\subfloat[]{\centering\includegraphics[width=.49\linewidth]{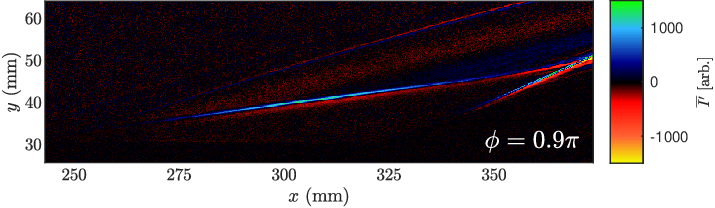}\label{25c2}}
\hfill
\subfloat[]{\centering\includegraphics[width=.49\linewidth]{Figures/Run25_phase0.5.eps}\label{25d2}}

\subfloat[]{\centering\includegraphics[width=.49\linewidth]{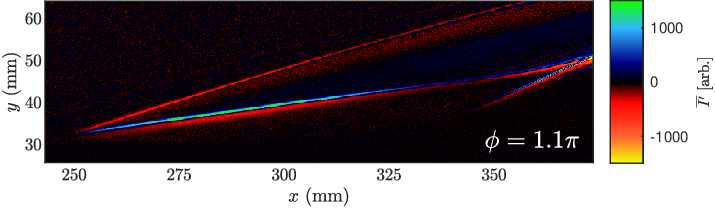}\label{25e2}}
\hfill
\subfloat[]{\centering\includegraphics[width=.49\linewidth]{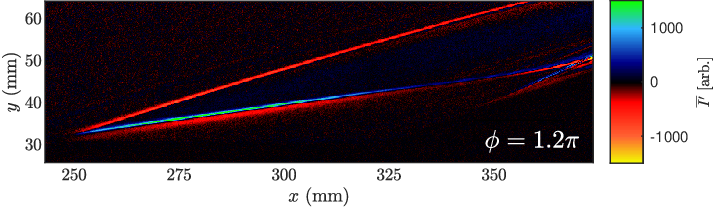}\label{25f2}}


\subfloat[]{\centering\includegraphics[scale=.75]{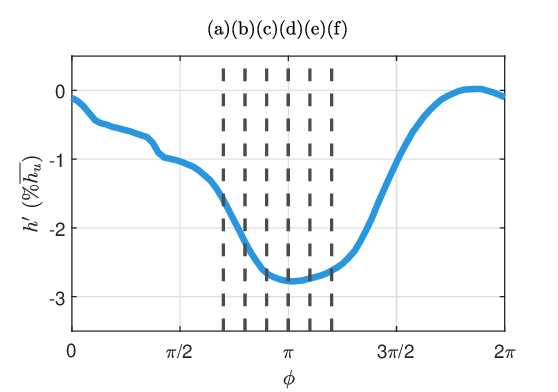}\label{25g2}}

 \caption{Mean-subtracted phase-averaged schlieren images showing intensity fluctuations during the post-disturbance deflection period for $f_p=$~1000~Hz. The separation-height phase average is included in (g) for reference, along with dashed lines indicating the corresponding phase for each image shown.}
 \label{PhAvSch_2}
 \end{figure}

Figure \ref{PhAvSch_2} shows phase-averaged schlieren intensity fluctuations for $f_p=$ 1000~Hz during the phases where the separation bubble is decreasing in height following the turbulent breakdown of the disturbances near reattachment. The decrease in separation height (blue region) moves upstream between $\phi=0.7 \pi -  \pi$, with the separation shock only following after the upstream-moving deflection has reached the shock foot. An interesting feature of these phase average images is the fainter region of decreased intensity (red) between the shock and the shear layer, which appears to travel upstream with the front of the shear layer deflection. The separation shock only begins to deflect once this red artifact reaches it, and the artifact does not move beyond the shock.  

\begin{figure}[h!]
 \includegraphics[width=0.6\linewidth]{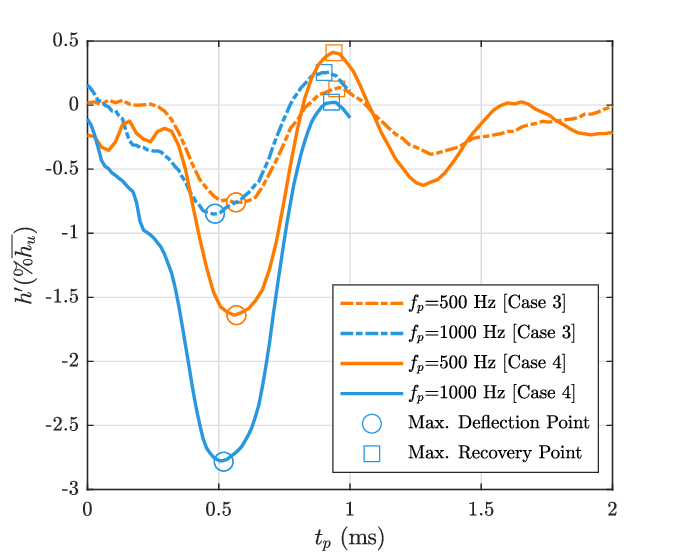}
 \centering
 \caption{Phase averages for $f_p$= 500 and 1000~Hz plotted with respect to time, with the maximum deflection and recovery points identified for each phase average [Cases 3 and 4]. }
 \label{Recovery}
 \end{figure}

As seen in figure \ref{Response}, the perturbation packets cause a temporary decrease in the separation bubble height after they pass through the shear layer. At 500~and~1000~Hz, the period between packets is long enough to observe how the separation bubble reacts once it has reached its minimum height. This provides insight into the separation bubble's response to a displacement from its mean position when the cause of the displacement is no longer present -- as the disturbances have convected off the back of the model by this point. Figure \ref{Recovery} shows the phase averages for these perturbation frequencies on a common time scale for the two highest-energy cases. In the period between $t_p$ = 0.45 and 1~ms, the separation bubble reaches its minimum height, or the point of maximum deflection, after which it rebounds up to a maximum recovery point; both of these points are identified in figure \ref{Recovery}. Quadratic fits are applied to the deflection and recovery regions to obtain greater precision in determining the location of the points of maximum deflection and recovery. The time period between these two points, hereafter referred to as the recovery time, $t_r$, is listed in table \ref{tab:table2} for each phase average in figure \ref{Recovery}. Between the points of maximum deflection and recovery, the phase average forms one half of a quasi-sinusoidal oscillation; the frequency corresponding to the full period is determined using the recovery times and is listed in table \ref{tab:table2} as the recovery frequency. The measured frequencies associated with the separation bubble's recovery all lie between 1203~and~1345~Hz. Notably, perturbation near these frequencies yielded the highest-amplitude oscillations in the separation bubble. Also, in the power spectra for the baseline case, the two highest energy peaks occurred at $f=$~1250~and~1350~Hz (see figure \ref{Run15:f3}). This is evidence that the separation bubble has a preferred oscillation frequency envelope which is related to the time scale of the restoring mechanism \citep{touber_low-order_2011}. A possible explanation for the observed behavior is that the traveling disturbances are increasing the fluid entrainment rate as they convect along the shear layer, temporarily decreasing the size of the bubble as a result; after which, the separation bubble recharges mass to complete the oscillation cycle. It is clear in table \ref{tab:table2} that the separation bubble requires a specific length of time to recharge. This would explain why the higher perturbation frequencies cause a more significant decrease in mean separation height, as seen in figure \ref{pre and post}, since the length of time between perturbations is shorter than the time scale of the recharge cycle.

 \begin{table}

\centering
\begin{tabular}{lcccccc}

Case& $f_p$, Hz& $t_r$, ms & Recovery Frequency ($1/2t_r$), Hz \\ \hline
3& 500&0.3838& 1303\\
4& 500&0.3717& 1345\\
3& 1000&0.4158&1203\\
4& 1000&0.4108&1217\\

\end{tabular}
\caption{\label{tab:table2} Recovery Dynamics of the Separation Bubble}
\end{table}

\begin{figure}[h!]
 \centering
 \subfloat[]{\includegraphics[width=0.5\linewidth]{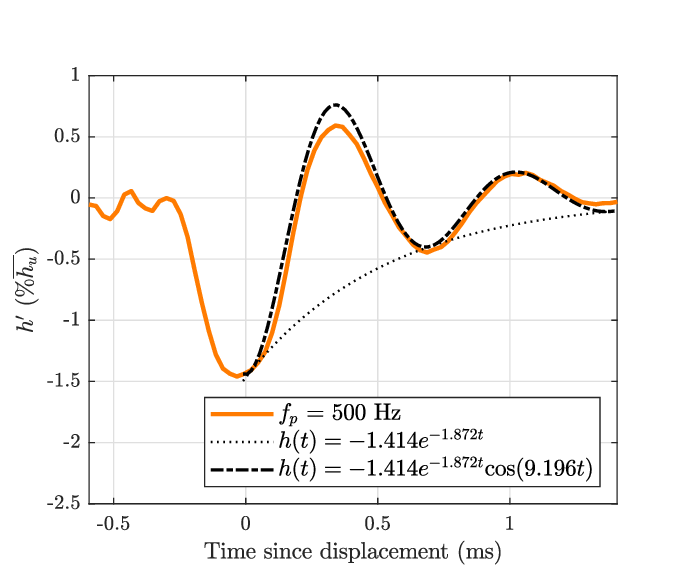}\label{dampA}}
  \hfill
      \subfloat[]{\includegraphics[width=0.5\linewidth]{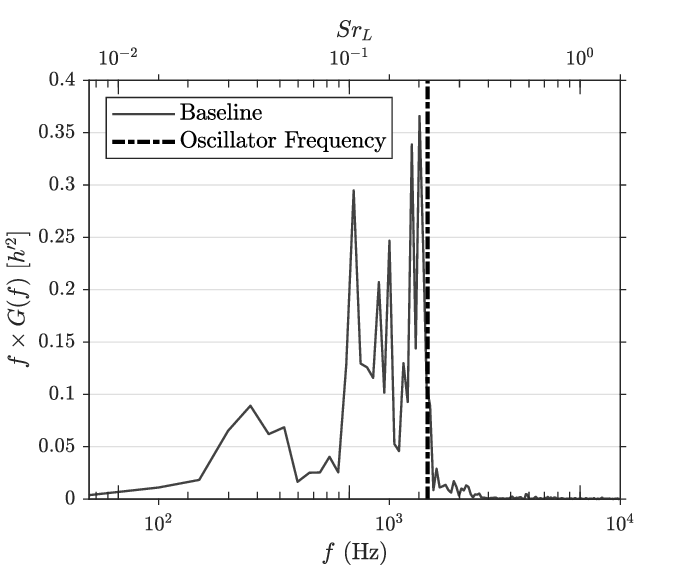}\label{dampB}}
      
    \subfloat[]{\includegraphics[width=0.5\linewidth]{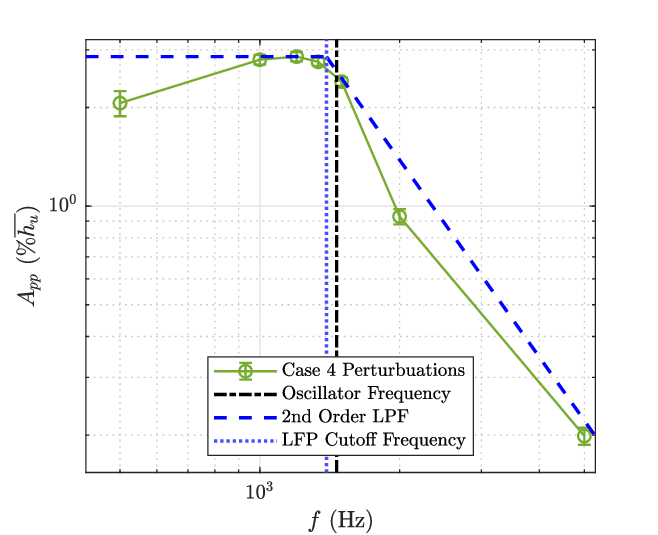}\label{dampC}}

 \caption{(a) Phase averaged separation-bubble fluctuations for $f_p = 500$ $\mathrm{Hz}$ (Case 4), with the displacement curve of an under-damped oscillator fit to the data. The y-axis has been offset so that the oscillations are centered around zero. (b) Pre-multiplied power spectral density plot of the baseline separation unsteadiness, with the dot-dashed line indicating the frequency of the under-damped oscillator equation. (c) Phase-average amplitudes for a range of perturbation frequencies plotted on a logarithmic scale alongside a 2nd-order low-pass filter model.}
 \label{Damping}
 \end{figure}

 It was proposed by \cite{clemens_low-frequency_2014} that the turbulent SWBLI acts as a forced dynamical system at low frequencies. This idea was supported by computations of laminar and turbulent interactions from \cite{bugeat_low-frequency_2022} and \cite{nichols_stability_2017}, respectively, with results from the latter suggesting that the system could be modeled as an under-damped oscillator. \cite{nicotra_perturbation_2026} experimentally observed a laminar axisymmetric separation bubble behaving similarly to an under-damped oscillator when one side of the shear layer was displaced from its stable position by a turbulent perturbation. In the present study, the spark-discharge-generated disturbances act as transient forcing resulting in a temporary displacement of the shear layer. Figure \ref{dampA} shows the 500~Hz phase average for Case 4 alongside an equation for the impulse response of an under-damped oscillator which has been fit to the data. The SWBLI system's response to this displacement closely resembles the displacement curve for an under-damped dynamical system. This is only visible when $f_p$ = 500~Hz, which is the one perturbation frequency with a period that is long enough to observe the system's extended response to the displacement. This observation strongly supports the theory that the interaction region is a weakly-damped dynamical system that responds to forcing from external disturbances \citep{porter_selective_2019,gaitonde_dynamics_2023}. This behavior is observed after the maximum displacement has been reached; by this time, the transient disturbances have convected off the back of the model and therefore are not influencing the system's response to this displacement. The equation for the under-damped displacement curve was determined by applying an exponential fit to the three local minima, and the oscillation period was determined from the time between the minima, in ms. The corresponding oscillator frequency is 1464~Hz, and the damping ratio, $\zeta$, is 0.199.  Introducing tailored forcing to excite an impulse response presents a technique to facilitate direct experimental measurements of the oscillator frequency and damping ratio of the dynamic system.
 
 
The SWBLI's unsteady response to broadband forcing has been frequently described as that of a low-pass filter or frequency-selective amplifier \citep{plotkin_shock_1975,touber_low-order_2011,poggie_spectral_2015,threadgill_transitional_2021,bugeat_low-frequency_2022}. Figure \ref{dampB} shows the baseline PSD with a vertical dot-dashed line representing the measured oscillation frequency of the under-damped system from figure \ref{dampA}. Interestingly, this frequency appears to be the approximate cutoff frequency of the system's natural unsteadiness.  Figure \ref{dampC} compares the response amplitudes at various perturbation frequencies with a 2nd-order low-pass filter model, along with the oscillator frequency (1464~Hz) and cutoff frequency of the model (1392~Hz). These results suggest that the SWBLI in this study can be modeled as an under-damped dynamic system with an oscillator frequency that acts as a cutoff frequency for the systems receptivity to external disturbances. This corroborates the theory proposed by \cite{bugeat_low-frequency_2022} in which the time-scale of the low-pass filter is set by the damping rate of the weakly-damped global mode. 

\subsection{Conclusion}

The dynamics of an interaction-transitional 3D-SWBLI are investigated by perturbing the system with disturbances generated by spark discharges. Experiments are performed in a Mach-6 quiet tunnel on a cone-slice-ramp model with a nanosecond-duration repetitively pulsed plasma actuator in the upstream boundary layer. The SWBLI unsteadiness is primarily characterized through high-speed schlieren imaging, using an algorithm to identify and track the separated shear layer. The algorithm's performance is validated against low-frequency static pressure fluctuations measured inside the separation bubble during the baseline case. Perturbation packets consist of ns-duration plasma pulses in groups of 5-10 at a pulse repetition frequency of 100~kHz. These disturbance packets are introduced into the upstream boundary layer at perturbation frequencies ranging from 500 to 5000~Hz. Four perturbation cases are tested, with pulse width and number varying with each case.

The baseline flow features a laminar upstream boundary layer, which separates at the leading edge of the slice, resulting in a fixed separation front. The unsteady motion of the SWBLI system consists of a large-scale flapping motion hinged at the separation point. Pressure fluctuations measured on the ramp show boundary-layer transition upon shear-layer reattachment. The SWBLI exhibits broadband separation unsteadiness at frequencies below 1500~Hz. There is a significant decline in the power content associated with frequencies beyond 1500~Hz, suggesting that this may be a natural cutoff frequency for the system. 

The vast majority of studies on SWBLI unsteadiness investigate the dynamic system under continuous forcing from turbulence in the boundary layer and freestream. In this study, transition upon reattachment is the only significant contributor to natural forcing, with the laminar upstream boundary layer and quiet freestream significantly reducing the magnitude of continuous forcing applied to the system. In this environment, tailored forcing is applied via spark-discharge-generated disturbances, which prompt turbulent breakdown in the downstream shear layer causing temporary excitation of the dynamic system. The measured separation fluctuations show that the disturbances in the boundary layer generally impose their frequency on the unsteady separation motion; however, multiple aspects of the separation response depend on perturbation frequency. When perturbing at frequencies above the natural cutoff frequency, there is a significant decrease in the magnitude of the induced oscillations at that frequency. Moreover, in the power spectra for these higher-frequency perturbations, there is a reduction in the energy corresponding to the natural oscillation frequencies when compared with the baseline case. 

Phase averages of the measured fluctuations provide insight into the mean effect of the perturbations on the separation bubble dynamics. Perturbation frequencies between 1000 and 1500~Hz yield the highest phase-averaged amplitudes, with a significant decrease in amplitude at perturbation frequencies beyond 1500~Hz. Increasing the energy deposited by a perturbation packet generally increases the amplitude of the phase-averaged oscillations. Comparisons between Cases 2 and 3, which have similar plasma packet energies but different pulse numbers and widths, reveals that the duration of the perturbation packet was a less important factor in exciting the low-frequency unsteadiness than the individual pulse energies. Considering that the disturbance packets are able to effectively excite low-frequency unsteadiness while the pulse and packet durations are one or more orders of magnitude below that of the unsteadiness, these results suggest that the duration of the disturbances does not need to be on the same time scale as the unsteadiness in order to yield a low-frequency response.

Analyses of the phase averaged separation fluctuations on a common time scale reveal that a consistent effect of the disturbances across perturbation frequencies is a delayed decrease in separation height. Phase averaged schlieren images show that this decrease in separation height is immediately preceded by the turbulent breakdown of the disturbances near reattachment. They also reveal that the displacement of the shear layer begins downstream and moves upstream and that this precedes the separation shock motion. Also, the average times required by the separation bubble to recover from displacement are measured, and the frequencies associated with this recovery time are between 1200 and 1345~Hz. Comparing these values to the perturbation frequencies that generate the highest-amplitude oscillations (1200 and 1335 Hz) and the highest-energy peaks in the power spectra of the baseline unsteadiness (1250 and 1350 Hz) suggests that the preferred oscillation frequency is tied to the time constant of the restoring mechanism. 

In the phase average for the lowest-frequency perturbation case, the period is long enough to observe the separation bubble's extended response to displacement past when the local maximum has been reached. The interaction's response strongly resembles the displacement curve of an under-damped oscillator with an associated frequency approximately equal to the cutoff frequency of the natural oscillations. These experimental findings support the theory that the SWBLI can be modeled as an under-damped dynamical system, and that separation unsteadiness arises from external forcing of this system. Furthermore, the similarity between the system cutoff frequency and the measured oscillator frequency provides evidence that the cutoff frequency is set by the damping rate of the weakly-damped global mode, which had been observed by \cite{bugeat_low-frequency_2022} using resolvent analysis.

\section*{Acknowledgments}
 The authors would like to thank Zachary McDaniel for the use of his SQUID program, which was used for data management and tunnel-condition analysis.

\section*{Funding Statement}
This work was supported in part by the United States Navy (NSWC-Dahlgren under contract number N00178-24-1-0005).

\section*{Competing Interests}
The authors declare none.

\bibliographystyle{jfm}
\bibliography{references1}

\end{document}